\pdfoutput=1
\documentclass[smallcondensed]{svjour3}    
\smartqed  
\usepackage{graphicx}
\usepackage{amssymb}
\usepackage{amsfonts}
\usepackage{enumerate} 
\usepackage{array}
\usepackage{amsmath}
\usepackage{comment}
\usepackage{color}
\usepackage{multirow}
\usepackage{numprint}

\usepackage{natbib}
\usepackage{url}

\usepackage{listings} 
\newcommand\crm{\cr\noalign{\medskip}}

\newcommand\norm[1]{\left\Vert#1\right\Vert}

\newcommand\be{\begin{equation}}
\newcommand\ee{\end{equation}}
\def\m@th{\mathsurround=0pt}
\newcommand\EQM[1]{\vcenter{\normalbaselines\m@th
    \ialign{${\displaystyle ##}$\hfil&&\ ${\displaystyle ##}$\hfil\crcr
    \mathstrut\crcr\noalign{\kern-\baselineskip}
    \noalign{\smallskip}
    #1\crcr\mathstrut\crcr\noalign{\kern-\baselineskip}}}}

\renewcommand\L{L}
\renewcommand\a{{\alpha}}

\newcommand\beq{\begin{equation}}
\newcommand\eeq{\end{equation}}

\newcommand\bvv{\begin{verbatim}}
\newcommand\evv{\end{verbatim}}

\renewcommand\a{{\alpha}}
\renewcommand\b{{\beta}}
\renewcommand\c{{\gamma}}
\renewcommand\d{{\delta}}

\newcommand{\bg}{\beta}

\journalname{Celestial Mechanics and Dynamical Astronomy}

\begin{document}

\title{Dedicated symplectic integrators for rotation motions
}


\author
{
	Jacques Laskar   \and
	Timoth\'ee Vaillant	
}


\institute{
	 Jacques Laskar \and
	Timoth\'ee Vaillant	\at
    ASD, IMCCE-CNRS UMR8028, Observatoire de Paris, PSL Universit\'{e}, Sorbonne Universit\'{e},
    77 Av. Denfert-Rochereau, 75014-Paris, France 
	\email{laskar@imcce.fr}
 }

\date{Received: \today/ Accepted: date}

\maketitle

\begin{abstract}

We propose to use the properties of the Lie algebra of the angular momentum to build symplectic integrators dedicated to the Hamiltonian of the free rigid body.
By introducing a dependence of the coefficients of integrators on the moments of inertia of the integrated body, we can construct symplectic dedicated integrators with fewer stages than in the general case for a splitting in three parts of the Hamiltonian.
We perform numerical tests to compare the developed dedicated 4th-order integrators to the existing reference integrators for the water molecule.
We also estimate analytically the accuracy of these new integrators for the set of the rigid bodies and conclude that they are more accurate than the existing ones only for very asymmetric bodies.

\end{abstract}

\renewcommand\L{L}

\section{Introduction}

The problem of the free rigid body is well-known to be integrable and the exact solution was developed by \cite{jacobi1850}. The solution uses Jacobi elliptic functions and elliptic integrals, which are necessary to solve exactly this problem. Nevertheless, the numerical evaluation of these functions is more expensive than the usual ones \citep[e.g.][]{toumawisdom1994,fasso2003}.
If only the final orientation of the body is necessary, only one step is required, and the higher cost is not a problem.
However in most cases, the kinetic energy of the free rigid body is coupled to a potential part making the problem no longer integrable.
It is notably the case in celestial mechanics for the integration of the dynamics of a planetary system where the rotation of a body interacts with the orbital motion \citep[e.g.][]{toumawisdom1994} and in molecular dynamics \citep[e.g.][]{dullweber1997}.
The temporary position of the body is then needed at every step and the use of a cheaper approximated integrator is significant. 

Several approximated integrators, which can be used for the free rigid body, exist (see \cite{hairer2006} for their description and \cite{hairervilmart2006} for a comparison). In this paper, we are only interested in the splitting technique. This symplectic method usually consists in the splitting of a Hamiltonian into several integrable parts, which are successively integrated by stage. It allows one to conserve on average the energy but it realizes the integration of a slightly perturbed Hamiltonian. Each stage of the integration scheme is weighted by a coefficient. The number of stages and an appropriate choice of their coefficients allow one to increase the order of the integrator.

This technique was proposed for the free rigid body by \cite{mclachlan1993}, \cite{toumawisdom1994}, \cite{reich1994} and splits the Hamiltonian into parts, which can be easily integrated as a succession of elementary rotations. The integration schemes of the free rigid body are those usually used for any Hamiltonian. For instance \cite{toumawisdom1994} and \cite{dullweber1997} used the classical St\"{o}rmer-Verlet or leapfrog integrator and \cite{omelyan2007} used the Yoshida's technique \citep{yoshida1990} to obtain a 4th-order scheme with the 2nd-order leapfrog scheme.
These integrators are then symmetrically composed with a potential to realize the symplectic integration of the perturbed rotation as it was done for instance by \cite{toumawisdom1994} in celestial mechanics and by \cite{dullweber1997} in molecular dynamics.

 \cite{fasso2003} compared the efficiency of the different possible splittings of the Hamiltonian for the 2nd-order leapfrog scheme by evaluating the 3rd-order remainder of the integrator, which dominates the error between the approximated integration and the exact solution. By computing the 3rd-order remainder for each scheme and each permutation of moments of inertia, \cite{fasso2003} concluded that the most efficient scheme depends on the moments of inertia of the considered body and particularly noticed that the Lie algebra of the angular momentum allows one to simplify the expression of the 3rd-order remainder.

The aim of this paper is to use the Lie algebra of the angular momentum to construct symplectic integrators dedicated to the Hamiltonian of the free rigid body more effective than the existing reference integrators. As noticed by \cite{fasso2003}, the number of terms in the 3rd-order remainder is lower than for an ordinary Hamiltonian. It is then possible to construct symplectic integrators with fewer stages. 
This is made possible by the fact that in the present work the coefficients of the integrator depend on the moments of inertia of the body. The determination of these coefficients is then allowed by the study of the Lie algebra of the angular momentum. Therefore each rigid body has its proper integrator with different coefficients.
In this paper, we are interested in developing symplectic integrators for the free rotation.
These integrators can be then coupled with a potential to integrate the perturbed rotation.

The structure of the remainder of a symplectic integrator of the free rigid body is developed in section \ref{sec:constraint}. In addition, relations between the coefficients of the remainder allow us to reduce the number of conditions on the integrator. In section \ref{sec:construction}, we construct symmetric integrators, which verify the conditions of section \ref{sec:constraint}. These integrators can have fewer stages than the usual integrators but are specific to a given rigid body because the coefficients of the integrators depend on their moments of inertia. We then proceed to numerical tests in section \ref{sec:numerical}: we first consider the simplest case of the spherical top and then the water molecule, which is an asymmetric body used in previous studies to test integrators of the rigid body. For these two selected bodies, we determine the best dedicated symplectic integrators and compare them to the usual symplectic integrators. In section \ref{sec:comparison}, we determine analytically the best dedicated symplectic integrators for the set of the rigid bodies and compare them to the usual schemes.

\section{Constraints on a symplectic integrator\label{sec:constraint}}

We consider a free rigid body $\mathcal{S}$ in the inertial reference frame $\mathcal{R}$. $(\mathbf{I},\mathbf{J},\mathbf{K})$ is a direct orthonormal basis associated with the body frame $\mathcal{R}_{S}$. The vectors $\mathbf{I}$, $\mathbf{J}$, $\mathbf{K}$ are associated with the principal axes of inertia of moments of inertia respectively $I_{1}$, $I_{2}$ and $I_{3}$. $\mathbf{g}$ is the angular momentum of $\mathcal{S}$ expressed in the inertial frame $\mathcal{R}$. For a free rigid body, the angular momentum $\mathbf{g}$ and the Hamiltonian $H$ are conserved. The Hamiltonian $H$ of the free rigid body is reduced to the rotational kinetic energy
\begin{equation}
H=\frac{G_{1}^{2}}{2I_{1}}+\frac{G_{2}^{2}}{2I_{2}}+\frac{G_{3}^{2}}{2I_{3}},
\end{equation}
where $\mathbf{G}=\left(G_{1},G_{2},G_{3}\right)$ is the angular momentum expressed in the body frame $\mathcal{R}_{S}$. In the body frame $\mathcal{R}_{S}$, the coordinates of $\mathbf{G}$ are not conserved but the norm $G$ is conserved with
\begin{equation}
G=\sqrt{G_{1}^{2}+G_{2}^{2}+G_{3}^{2}}.
\end{equation}

Two different splittings of the Hamiltonian $H$ can reduce its integration to the composition of simple rotations. The first splitting, ABC, splits the Hamiltonian into three parts \citep[e.g.][]{reich1994}
\begin{equation}
H=A+B+C,
\end{equation} 
with
\begin{equation}
A=\frac{G_{1}^{2}}{2I_{1}},
\end{equation}
\begin{equation}
B=\frac{G_{2}^{2}}{2I_{2}},
\end{equation}
\begin{equation}
C=\frac{G_{3}^{2}}{2I_{3}}.
\end{equation} 
For the length of time $t$, each part corresponds to a rotation of angles $G_{1}t/I_{1}$, $G_{2}t/I_{2}$, $G_{3}t/I_{3}$, around the respective principal axes $\mathbf{I}$, $\mathbf{J}$, $\mathbf{K}$. The second splitting, RS, splits the Hamiltonian into two parts \citep{mclachlan1993,toumawisdom1994}
\begin{equation}
H=R+S,
\end{equation} 
with
\begin{equation}
R=\frac{G_{1}^{2}}{2}\left(\frac{1}{I_{1}}-\frac{1}{I_{2}}\right),\label{eq:step_R}
\end{equation}
\begin{equation}
S=\frac{G_{3}^{2}}{2}\left(\frac{1}{I_{3}}-\frac{1}{I_{2}}\right)+\frac{G^{2}}{2I_{2}}.\label{eq:step_S}
\end{equation}  
$R$ corresponds to the rotation around the principal axis $\mathbf{I}$ with the angle $G_{1}t(1/I_{1}-1/I_{2})$ and $S$, which is the Hamiltonian of a symmetric top, to a rotation around the principal axis $\mathbf{K}$ of angle $G_{3}t(1/I_{3}-1/I_{2})$ followed by a rotation around the angular momentum $\mathbf{G}$ of angle $Gt/I_{2}$. 

These two decompositions give rise to two possible splittings, which result in two classes of symplectic integrators,
\begin{equation}
\mathcal{S}_{ABC}\left(h\right)=\prod_{i=1}^{n} e^{a_{i}hL_{A}}e^{b_{i}hL_{B}}e^{c_{i}hL_{C}},\label{eq:scheme_ABC}
\end{equation}
and
\begin{equation}
\mathcal{S}_{RS}\left(h\right)=\prod_{i=1}^{n} e^{a_{i}hL_{R}}e^{b_{i}hL_{S}},\label{eq:scheme_RS}
\end{equation}
where $L_{X}= \left\lbrace X,.\right\rbrace$ is the Lie derivative of a Hamiltonian $X$, $h$ the step size and $a_i$, $b_i$ and $c_i$ the coefficients of the integrators.

These splitting integrators exactly integrate a slightly different Hamiltonian $K$. For the symplectic integrator $\mathcal{S}\left(h\right)=e^{hL_K}$, this Hamiltonian is given by \citep[e.g.][]{yoshida1990,koseleff1993}
\begin{equation}
hK=hH+\sum^{n}_{k=2} h^{k}H_{R_{k}}+O\left(h^{n+1}\right),
\end{equation}
where each Hamiltonian $H_{R_{k}}$ is the remainder of order $k$.

The remainders $H_{R_k}$ of the schemes $\mathcal{S}_{ABC}$ and $\mathcal{S}_{RS}$ belong to the Lie algebra $\mathcal{L}$ generated by the alphabet $\mathcal{A}$ composed of the three elements $G^2_1$, $G^2_2$, $G^2_3$ and associated with the Poisson brackets. 
$\mathcal{L}=\oplus_{k\geq 1} \mathcal{L}_k$ is a graded Lie algebra and is the sum of the Lie algebras $\mathcal{L}_k$ generated by the Lie monomials of length $k$ \citep[e.g.][]{koseleff1993}.
The Hamiltonian $H_{R_{k}}$ belongs to $\mathcal{L}_k$ and is the sum of Lie monomials of length $k$.

To obtain an integrator of order $n$, we must have $H_{R_{k}}=0$ for $k=2,\ldots,n$ \citep[e.g.][]{yoshida1990,koseleff1993,mclachlan1995}. 
The Baker-Campbell-Hausdorff formula allows one to determine the remainders for each order and to know what equations must verify the coefficients $a_i$, $b_i$ and $c_i$ to cancel the remainders $H_{R_{k}}$ for $k=2,\ldots,n$.
If the scheme is symmetric, $H_{R_{k}}=0$ is already verified for the even values of $k$ \citep[e.g.][]{yoshida1990}.

The number of independent equations at order $k$ which must verify the coefficients $a_i$, $b_i$ and $c_i$ to verify $H_{R_{k}}=0$ is given by the dimension of the Lie algebra $\mathcal{L}_k$.
The minimal number of stages of an integrator of order $n$ is then given by the total number of independent equations, which the coefficients $a_i$, $b_i$ and $c_i$ must verify to have an integrator of order $n$.
  
\cite{fasso2003} compared the efficiency of the two possible splittings for the Hamiltonian of the free rigid body for the 2nd-order symmetric schemes obtained with the leapfrog method
\begin{eqnarray}
\mathcal{S}_{ABCBA2}\left(h\right) & = & e^{\frac{h}{2}L_{A}}e^{\frac{h}{2}L_{B}}e^{hL_{C}}e^{\frac{h}{2}L_{B}}e^{\frac{h}{2}L_{A}}, \\
\mathcal{S}_{RSR2}\left(h\right) & = & e^{\frac{h}{2}L_{R}}e^{hL_{S}}e^{\frac{h}{2}L_{R}}, \\
\mathcal{S}_{SRS2}\left(h\right) & = & e^{\frac{h}{2}L_{S}}e^{hL_{R}}e^{\frac{h}{2}L_{S}}.
\end{eqnarray}
To obtain all the possible schemes, \cite{fasso2003} considered the six permutations of the three Hamiltonians $G_1^2/(2I_1)$, $G_2^2/(2I_2)$ and $G_3^2/(2I_3)$, which is equivalent to consider the permutations of the moments of inertia.
We call the six permutations $ABC$, $BCA$, $CAB$, $ACB$, $CBA$, $BAC$.

For each scheme, \cite{fasso2003} simplified the analytical expression of the three order remainder by using the relation $\lbrace G_{i},G_{j} \rbrace = \epsilon_{ijk} G_{k}$ for the Poisson brackets $\lbrace , \rbrace$ and estimated it for the six permutations. 
He concluded that their efficiency depends on the moments of inertia for a given body. 
For the bodies near to a symmetric top, the integrators $\mathcal{S}_{RSR2}$ and $\mathcal{S}_{SRS2}$ are more accurate than $\mathcal{S}_{ABCBA2}$. 
It is possible to combine symmetrically these three second order integrators to obtain higher order integrators  \citep[e.g.][]{suzuki1990,yoshida1990,mclachlan1995}. 
These techniques work for any Hamiltonian but do not consider the Lie algebra of the free rigid body.

In section \ref{sec:construction}, we seek to construct symmetric 4th-order integrators for the Hamiltonian of the free rigid body.
To obtain a 4th-order integrator, we must verify $H_{R_2}=H_{R_3}=H_{R_4}=0$.
For a symmetric integrator, the remainders of even order are already canceled and we must then just verify $H_{R_3}=0$.
To know the number of coefficients necessary to cancel this remainder, we need to know its expression.

In this section, we then study the structure of the Lie algebra for the Hamiltonian of the free rigid body to know the number of constraints to impose to construct symplectic integrators for the free rigid body.

\subsection{Lie algebra structure for the Hamiltonian of the free rigid body\label{sec:constraint_algebra}}

The elements of the Lie algebra $\mathcal{L}_k$ are the sum of Lie monomials of length $k$ for the alphabet $\mathcal{A}=(G_1^2,G_2^2,G_3^2)$.
The Poisson brackets of the components $G_i$ of the angular momentum verify the relation \citep[e.g.][]{toumawisdom1994,fasso2003}
\begin{equation}
\left\lbrace G_{i},G_{j} \right\rbrace = \epsilon_{ijk} G_{k}.\label{eq:Lie}
\end{equation}
Here we take into account this relation to express the elements of the Lie algebra $\mathcal{L}_k$ of order $k$ as a linear combination of monomials of the components $G_i$ of the angular momentum.

\subsubsection{First orders\label{sec:constraint_first}}

We first look the structure of the algebra for the first orders.

If the family of elements $v_{1_i}$ spans $\mathcal{L}_{1}$ and the family of elements $v_{k_j}$ spans $\mathcal{L}_{k}$, the family of elements $\lbrace v_{1_i},v_{k_j} \rbrace$ spans $\mathcal{L}_{k+1}$.
To obtain the expression of an element of $\mathcal{L}_{k+1}$, we must then compute all the terms $\lbrace v_{1_i},v_{k_j}\rbrace$.
We note $n_k$ the number of monomials in the linear combination for the order $k$.

\paragraph{Order 1:}
For the first order, the Lie monomials of length 1 are $G_{1}^{2}$, $G_{2}^{2}$, $G_{3}^{2}$. Therefore $n_1=3$.  
\paragraph{Order 2:}
For the second order, the Lie monomials of length 2 can be expressed as $\lbrace  G_{j}^{2},G_{k}^{2}\rbrace$. With Eq. (\ref{eq:Lie}), we have 
\begin{equation}
\left\lbrace G_{1}^2,G_{2}^2\right\rbrace = \left\lbrace G_{2}^2,G_{3}^2\right\rbrace = \left\lbrace G_{3}^2,G_{1}^2\right\rbrace = 4G_{1}G_{2}G_{3},
\end{equation}
and therefore $n_2=1$.
($G_1G_2G_3$) is a basis of $\mathcal{L}_2$ and then the dimension of $\mathcal{L}_2$ is $1$.
To simplify, we note $W=G_{1}G_{2}G_{3}$ in the following.

\paragraph{Order 3:}
For the third order, the Lie monomials of length 3 can be written as $\lbrace G_{i}^{2},\lbrace G_{j}^{2},G_{k}^{2}\rbrace\rbrace$ and are a linear combination of the terms $\lbrace G_{i}^{2},W\rbrace$. With Eq. (\ref{eq:Lie}), we have
\begin{eqnarray}
\left\lbrace G_{1}^2,W\right\rbrace & = & 2\left(G_{1}^2G_{3}^2-G_{1}^2G_{2}^2\right)  \\
\left\lbrace G_{2}^2,W\right\rbrace & = & 2\left(G_{2}^2G_{1}^2-G_{2}^2G_{3}^2\right)  \\
\left\lbrace G_{3}^2,W\right\rbrace & = & 2\left(G_{3}^2G_{2}^2-G_{3}^2G_{1}^2\right).
\end{eqnarray}
Therefore, each element of $\mathcal{L}_3$ is a linear combination of the terms $G_{1}^{2}G_{2}^{2}$, $G_{1}^{2}G_{3}^{2}$, $G_{2}^{2}G_{3}^{2}$ and $n_3=3$. 

We consider the 3rd-order remainder $H_{R_{3}}$ of an integrator of the free rigid body. $H_{R_{3}}$ belongs to $\mathcal{L}_{3}$ and is then a linear combination of the three terms $G_{1}^2G_{3}^2-G_{1}^2G_{2}^2$, $G_{2}^2G_{1}^2-G_{2}^2G_{3}^2$, $G_{3}^2G_{2}^2-G_{3}^2G_{1}^2$, 
\begin{equation}
H_{R_{3}}\left(G_{1},G_{2},G_{3}\right) =P_1 G_{1}^{2}G_{2}^{2} + P_2 G_{1}^{2}G_{3}^{2} + P_3 G_{2}^{2}G_{3}^{2}
\end{equation} 
with the coefficients $P_{i}$. If $G_{1}=G_{2}=G_{3}=1$, $G_{1}^2G_{3}^2-G_{1}^2G_{2}^2$, $G_{2}^2G_{1}^2-G_{2}^2G_{3}^2$, $G_{3}^2G_{2}^2-G_{3}^2G_{1}^2$ are canceled and $H_{R_{3}}\left(1,1,1\right)=0$. The coefficients $P_{i}$ must then verify
\begin{equation}
P_1 + P_2 + P_3=0. \label{eq:P_k=0}
\end{equation}
The three terms are linearly dependent. By keeping two of them, we can easily verify that we obtain two linearly independent terms. For instance, ($G_{1}^2G_{3}^2-G_{1}^2G_{2}^2$, $G_{2}^2G_{1}^2-G_{2}^2G_{3}^2$) is a basis of $\mathcal{L}_{3}$ and the dimension of $\mathcal{L}_{3}$ is then $2$.

\paragraph{Orders 4 and 5:}We can continue this procedure for the orders 4 and 5. We obtain that ($G_1^2W$, $G_2^2W$, $G_3^2W$) is a basis of $\mathcal{L}_4$ of dimension $3$ and that ($3W^2-G_1^4G_2^2$, $3W^2-G_2^4G_3^2$, $3W^2-G_3^4G_1^2$, $G_1^4(G_2^2-G_3^2)$, $G_2^4(G_3^2-G_1^2)$, $G_3^4(G_1^2-G_2^2)$) is a basis of $\mathcal{L}_5$ of dimension $6$.

\subsubsection{All orders}

We can generalize the structure of the Lie algebras for the first orders with a general theorem about the Lie algebra generated by the alphabet $\mathcal{A}=(G_1^2,G_2^2,G_3^2)$.

\paragraph{Theorem 1:}

Let $\mathcal{L}$ be the graded Lie algebra generated by the alphabet $\mathcal{A}=(G_1^2,G_2^2,G_3^2)$ with $G_i$ the components of an angular momentum and $\mathcal{L}_k$ the Lie algebra generated by the Lie monomials of length $k$ such that $\mathcal{L}=\oplus_{k\geq 1} \mathcal{L}_k$. 

Each element of $\mathcal{L}_{2k}$ for $k\in\mathbb{N}^*$ can be expressed as a linear combination of the monomials of degree $2k+1$
\begin{equation}
G_1^{2p+1}G_2^{2q+1}G_3^{2r+1}
\end{equation}
with $p+q+r=k-1$ and $p,q,r\in \mathbb{N}$.

Each element of $\mathcal{L}_{2k+1}$ for $k\in\mathbb{N}^*$ can be expressed as a linear combination of the monomials of degree $2k+2$
\begin{equation}
G_1^{2p}G_2^{2q}G_3^{2r}
\end{equation}
with $p+q+r=k+1$, $p,q,r\in \mathbb{N}$ and $p,q,r\leq k$.

\paragraph{Demonstration:}
We proceed by recurrence to demonstrate the theorem 1 and consider the proposition $\mathcal{P}_{k}$: each element of $\mathcal{L}_{2k}$ is a linear combination of the monomials $G_1^{2p+1}G_2^{2q+1}G_3^{2r+1}$ with $p+q+r=k-1$ and $p,q,r\in \mathbb{N}$.

For the order 2, we have seen in section \ref{sec:constraint_first} that each element of the Lie algebra $\mathcal{L}_2$ is proportional to $G_{1}G_{2}G_{3}$ and $\mathcal{P}_{1}$ is then true.

We suppose that $\mathcal{P}_{k}$ is true.
We know the family $(G_{1}^{2}, G_{2}^{2}, G_{3}^{2})$ which spans $\mathcal{L}_{1}$ and the family $(G_1^{2p+1}G_2^{2q+1}G_3^{2r+1})$ which spans $\mathcal{L}_{2k}$.
Each element of $\mathcal{L}_{2k+1}$ can then be written as a linear combination of the terms $\lbrace G_i^{2},G_1^{2p+1}G_2^{2q+1}G_3^{2r+1} \rbrace$.
We compute all these terms and for instance, we have
\begin{eqnarray}
& & \left\lbrace G_1^{2},G_1^{2p+1}G_2^{2q+1}G_3^{2r+1} \right\rbrace \nonumber \\
& = & G_1^{2p+1}G_3^{2r+1}\left\lbrace G_1^{2},G_2^{2q+1} \right\rbrace + G_1^{2p+1}G_2^{2q+1}\left\lbrace G_1^{2},G_3^{2r+1} \right\rbrace \nonumber \\
& = & 2\left(2q+1\right)G_1^{2\left(p+1\right)}G_2^{2q}G_3^{2r+1}\left\lbrace G_1,G_2 \right\rbrace - 2\left(2r+1\right)G_1^{2\left(p+1\right)}G_2^{2q+1}G_3^{2r}\left\lbrace G_1,G_3 \right\rbrace \nonumber \\
& = & \left(4q+2\right)G_1^{2\left(p+1\right)}G_2^{2q}G_3^{2\left(r+1\right)}-\left(4r+2\right)G_1^{2\left(p+1\right)}G_2^{2\left(q+1\right)}G_3^{2r}.
\end{eqnarray}
We have then 
\begin{equation}
\begin{array}{lll}
\left\lbrace G_1^{2},G_1^{2p+1}G_2^{2q+1}G_3^{2r+1} \right\rbrace = & & \left(4q+2\right)G_1^{2\left(p+1\right)}G_2^{2q}G_3^{2\left(r+1\right)} \\
 & - & \left(4r+2\right)G_1^{2\left(p+1\right)}G_2^{2\left(q+1\right)}G_3^{2r} \\
\left\lbrace G_2^{2},G_1^{2p+1}G_2^{2q+1}G_3^{2r+1} \right\rbrace = & & \left(4r+2\right)G_1^{2\left(p+1\right)}G_2^{2\left(q+1\right)}G_3^{2r} \\
& - & \left(4p+2\right)G_1^{2p}G_2^{2\left(q+1\right)}G_3^{2\left(r+1\right)} \\
\left\lbrace G_3^{2},G_1^{2p+1}G_2^{2q+1}G_3^{2r+1} \right\rbrace = & & \left(4p+2\right)G_1^{2p}G_2^{2\left(q+1\right)}G_3^{2\left(r+1\right)} \\
& - & \left(4q+2\right)G_1^{2\left(p+1\right)}G_2^{2q}G_3^{2\left(r+1\right)}. \\
\end{array} \label{eq:rel2k_2k+1}
\end{equation} 
We deduce that each element of $\mathcal{L}_{2k+1}$ is a linear combination of the monomials $G_1^{2p}G_2^{2q}G_3^{2r}$ with $p+q+r=k+1$, $p,q,r\in \mathbb{N}$ and $p,q,r\leq k$. Therefore, if the theorem 1 is verified for the order $2k$, it is also satisfied for the order $2k+1$.

Therefore, each element of $\mathcal{L}_{2k+2}$ can be written as a linear combination of the terms $\lbrace G_i^2,G_1^{2p}G_2^{2q}G_3^{2r} \rbrace$ with $p+q+r=k+1$, $p,q,r\in\mathbb{N}$ and $p,q,r\leq k$.
We then compute these terms
\begin{equation}
\begin{array}{lll}
\left\lbrace G_1^{2},G_1^{2p}G_2^{2q}G_3^{2r} \right\rbrace & = & 4qG_1^{2p}G_2^{2\left(q-1\right)}G_3^{2r}W-4rG_1^{2p}G_2^{2q}G_3^{2\left(r-1\right)}W \\
\left\lbrace G_2^{2},G_1^{2p}G_2^{2q}G_3^{2r} \right\rbrace & = & 4rG_1^{2p}G_2^{2q}G_3^{2\left(r-1\right)}W-4pG_1^{2\left(p-1\right)}G_2^{2q}G_3^{2r}W \\
\left\lbrace G_3^{2},G_1^{2p}G_2^{2q}G_3^{2r} \right\rbrace & = & 4pG_1^{2\left(p-1\right)}G_2^{2q}G_3^{2r}W-4qG_1^{2p}G_2^{2\left(q-1\right)}G_3^{2r}W. \\
\end{array} \label{eq:rel2k+1_2k+2}
\end{equation} 
We deduce that each element of $\mathcal{L}_{2k+2}$ is a linear combination of the monomials $G_1^{2p+1}G_2^{2q+1}G_3^{2r+1}$ with $p+q+r=k$ and $p,q,r\in\mathbb{N}$. If the theorem 1 is verified for the order $2k+1$, it is also satisfied for the order $2k+2$. 

If the proposition $\mathcal{P}_k$ is true, $\mathcal{P}_{k+1}$ is also verified. Therefore the theorem 1 is verified for all the integers $k$ with $k\geq 2$.

\subsection{Reduction formula}

For the 3rd-order, we have obtained the supplementary relation Eq. (\ref{eq:P_k=0}) between the coefficients of the 3rd-order remainder $H_{R_3}$. This allows us to decrease the number of independent coefficients needed to cancel $H_{R_3}$. This reduction formula can be generalized at all order with the following theorem. 

\paragraph{Theorem 2:}

Let $\mathcal{L}$ be the graded Lie algebra generated by the alphabet $\mathcal{A}=(G_1^2,G_2^2,G_3^2)$ with $G_i$ the components of an angular momentum and $\mathcal{L}_k$ the Lie algebra generated by the Lie monomials of length $k$ such that $\mathcal{L}=\oplus_{k\geq 1} \mathcal{L}_k$. Let $X\in \mathcal{L}_{2k+1}$ and $X=\sum_{\substack{0\leq p,q,r\leq k \\ p+q+r=k+1}} \beta_{2k+1,pqr} G_1^{2p}G_2^{2q}G_3^{2r}$.

The coefficients $\beta_{2k+1,pqr}$ verify the reduction formula
\begin{equation}
\sum_{\substack{0\leq p,q,r\leq k \\ p+q+r=k+1}} \frac{\left(2p\right)!}{2^p p!}\frac{\left(2q\right)!}{2^q q!}\frac{\left(2r\right)!}{2^r r!}\beta_{2k+1,pqr}=0.\label{eq:coef=0}
\end{equation}

\paragraph{Demonstration:}
We seek the coefficients $\lambda_{pqr}\neq0$ which satisfy the relation 
\begin{equation}
\sum_{\substack{0\leq p,q,r\leq k \\ p+q+r=k+1}} \lambda_{pqr} \beta_{2k+1,pqr}=0.
\end{equation}

Along the theorem 1, each element $X$ of $\mathcal{L}_{2k+1}$ can be written \begin{equation}
X=\sum_{\substack{0\leq p,q,r\leq k \\ p+q+r=k+1}} \beta_{2k+1,pqr} G_1^{2p}G_2^{2q}G_3^{2r}.
\end{equation}
Along the demonstration of the theorem 1, $X$ can also be written as
\begin{equation}
X=\sum^{3}_{i=1}  \sum_{\substack{0\leq p,q,r \\ p+q+r=k-1}} \alpha_{i,pqr} \left\lbrace G_i^{2},G_1^{2p+1}G_2^{2q+1}G_3^{2r+1} \right\rbrace,
\end{equation}
with
\begin{equation}
\begin{array}{lll}
\beta_{2k+1,pqr} & = & \left(4p+2\right)\left(\alpha_{3,pq-1r-1}-\alpha_{2,pq-1r-1}\right) \\
& + & \left(4q+2\right)\left(\alpha_{1,p-1qr-1}-\alpha_{3,p-1qr-1}\right) \\
& + & \left(4r+2\right)\left(\alpha_{2,p-1q-1r}-\alpha_{1,p-1q-1r}\right), \\
\end{array} \label{eq:beta_alpha}
\end{equation}
where $\alpha_{i,pqr}=0$ if $p$, $q$, $r$ do not verify $0\leq p,q,r$ and $p+q+r=k-1$.
From Eq. (\ref{eq:beta_alpha}), we deduce 
\begin{eqnarray}
 &  & \sum_{\substack{0\leq p,q,r\leq k \\ p+q+r=k+1}} \lambda_{pqr} \beta_{2k+1,pqr} \\ \nonumber
= & & \sum_{\substack{0\leq p,q,r\leq k \\ p+q+r=k+1}} \lambda_{pqr} \left(4p+2\right)\left(\alpha_{3,pq-1r-1}-\alpha_{2,pq-1r-1}\right) \\ \nonumber
& + & \sum_{\substack{0\leq p,q,r\leq k \\ p+q+r=k+1}} \lambda_{pqr} \left(4q+2\right)\left(\alpha_{1,p-1qr-1}-\alpha_{3,p-1qr-1}\right) \\ \nonumber
& + & \sum_{\substack{0\leq p,q,r\leq k \\ p+q+r=k+1}} \lambda_{pqr} \left(4r+2\right)\left(\alpha_{2,p-1q-1r}-\alpha_{1,p-1q-1r}\right) \\ \nonumber
= & & \sum_{\substack{0\leq p,q,r\leq k-1 \\ p+q+r=k-1}} \lambda_{pq+1r+1} \left(4p+2\right)\left(\alpha_{3,pqr}-\alpha_{2,pqr}\right) \\ \nonumber
& + & \sum_{\substack{0\leq p,q,r\leq k-1 \\ p+q+r=k-1}} \lambda_{p+1qr+1} \left(4q+2\right)\left(\alpha_{1,pqr}-\alpha_{3,pqr}\right) \\ \nonumber
& + & \sum_{\substack{0\leq p,q,r\leq k-1 \\ p+q+r=k-1}} \lambda_{p+1q+1r} \left(4r+2\right)\left(\alpha_{2,pqr}-\alpha_{1,pqr}\right) \\ \nonumber
= & & \sum_{\substack{0\leq p,q,r\leq k-1 \\ p+q+r=k-1}} \left( \lambda_{p+1qr+1} \left(4q+2\right)-\lambda_{p+1q+1r} \left(4r+2\right)\right) \alpha_{1,pqr} \\ \nonumber
& + & \sum_{\substack{0\leq p,q,r\leq k-1 \\ p+q+r=k-1}} \left( \lambda_{p+1q+1r} \left(4r+2\right)-\lambda_{pq+1r+1} \left(4p+2\right)\right)\alpha_{2,pqr} \\ \nonumber
& + & \sum_{\substack{0\leq p,q,r\leq k-1 \\ p+q+r=k-1}} \left( \lambda_{pq+1r+1} \left(4p+2\right)-\lambda_{p+1qr+1} \left(4q+2\right) \right) \alpha_{3,pqr}, \\ \nonumber
\end{eqnarray}
because $\alpha_{i,pqr}=0$ if $p$, $q$, $r$ do not verify $0\leq p,q,r$ and $p+q+r=k-1$. To verify Eq. (\ref{eq:coef=0}), it is sufficient to have
\begin{equation}
\lambda_{pq+1r+1} \left(2p+1\right)=\lambda_{p+1qr+1}\left(2q+1\right)=\lambda_{p+1q+1r}\left(2r+1\right). \label{eq:lambda_pqr}
\end{equation}  
To verify this relation, it is sufficient to have
\begin{equation}
\lambda_{pqr}=\frac{\left(2p\right)!}{2^p p!}\frac{\left(2q\right)!}{2^q q!}\frac{\left(2r\right)!}{2^r r!}.
\end{equation} 
Eq. (\ref{eq:coef=0}) allows us to find in an other way Eq. (\ref{eq:P_k=0}) obtained previously. 

\subsection{Number of stages of a symplectic integrator}

We consider the two possible integrators $\mathcal{S}_{ABC}$ (Eq. (\ref{eq:scheme_ABC})) and $\mathcal{S}_{RS}$ (Eq. (\ref{eq:scheme_RS})).
From the theorems 1 and 2, the expression of the modified Hamiltonian of these integrators is 
\begin{eqnarray}
hK & = & h\left( \bg_{1,100}\frac{G_1^{2}}{2I_1}+\bg_{1,010}\frac{G_2^{2}}{2I_2}+\bg_{1,001}\frac{G_3^{2}}{2I_3} \right) \label{eq:secspec_bcoef}\\ \nonumber
& + & \sum^{+\infty}_{k=1}h^{2k}\sum_{\substack{0\leq p,q,r \\ p+q+r=k-1}}\bg_{2k,pqr}G_1^{2p+1}G_2^{2q+1}G_3^{2r+1}\\ \nonumber
& + & \sum^{+\infty}_{k=1}h^{2k+1}\sum_{\substack{0\leq p,q,r\leq k \\ p+q+r=k+1}}\bg_{2k+1,pqr}G_1^{2p}G_2^{2q}G_3^{2r}, 
\end{eqnarray}
where the coefficients $\bg_{k,pqr}$ depend on the coefficients of the integrator $a_i$, $b_i$, $c_i$ and of the moments of inertia $I_1$, $I_2$, $I_3$.

To obtain an integrator of order $n$, it is sufficient to cancel all the coefficients $\b_{k,pqr}$ for $k\in\mathbb{N}$, $2\leq k \leq n$ and we must also verify $\b_{1,100}=\b_{1,010}=\b_{1,001}=1$.
We have in total $N\left(n\right)$ equations to verify at order $n$.
To solve $N\left(n\right)$ independent equations, we need $N\left(n\right)$ independent variables $a_i$, $b_i$, $c_i$ \citep[e.g.][]{koseleff1993,koseleff1996,mclachlan1995}. The number of stages of the integrator is then given by the number of equations. 

For the special case of a symmetric integrator, the remainders of even order are already canceled \citep[e.g.][]{yoshida1990} and the expression of the modified Hamiltonian is
\begin{eqnarray}
hK & = & h\left( \bg_{1,100}\frac{G_1^{2}}{2I_1}+\bg_{1,010}\frac{G_2^{2}}{2I_2}+\bg_{1,001}\frac{G_3^{2}}{2I_3} \right) \\ \nonumber
& + & \sum^{+\infty}_{k=1}h^{2k+1}\sum_{\substack{0\leq p,q,r\leq k \\ p+q+r=k+1}}\bg_{2k+1,pqr}G_1^{2p}G_2^{2q}G_3^{2r}.
\end{eqnarray}
To obtain an integrator of order $2n$, it is sufficient to cancel all the coefficients $\b_{2k+1,pqr}$ for $k\in\mathbb{N}$, $1\leq k \leq n-1$ and to verify $\b_{1,100}=\b_{1,010}=\b_{1,001}=1$. We have in total $N\left(2n\right)$ equations and the number of independent variables $a_i$, $b_i$, $c_i$ is then $N\left(2n\right)$. For a symmetric integrator of $2l$ stages, the stages $i$ and $2l-i$ are identical and the number of independent variables is $l$. Therefore the number of stages of the integrator is $2N\left(2n\right)$. However the stages $l$ and $l+1$ are identical and consecutive, and the number of stages becomes then $2N\left(2n\right)-1$.

\subsection{Application to the rigid body\label{sec:constraint_number}}

We now count the number of equations to obtain the number of stages of the integrators for the Hamiltonian of the free rigid body.
The number of independent equations for each order $k$ is given by the dimension of the Lie algebra $\mathcal{L}_k$.
However in section \ref{sec:constraint_algebra}, we have only determined the dimensions of the Lie algebras $\mathcal{L}_k$ for the first orders $k=1,2,3,4,5$.
We can then know the minimal number of stages only for these five first orders.
For the higher orders, we have only express the remainders as a linear combination of monomials and we obtain in this case an upper bound for the minimal number of stages.

\newcolumntype{P}[1]{>{\raggedleft}p{#1}}

\begin{table}
\begin{minipage}{1\textwidth}
\centering
\begin{tabular}{|P{0.8cm}|P{1.3cm}|*{6}{P{1.05cm}|} }
\hline
\multirow{3}{*}{Order} & \multirow{3}{*}{Dimension} & \multicolumn{4}{c|}{Number of equations} & \multicolumn{2}{c|}{Number of stages} \tabularnewline
\cline{3-8}
	  & & \multicolumn{2}{c|}{NS} & \multicolumn{2}{c|}{S} & \multicolumn{1}{c|}{NS} & \multicolumn{1}{c|}{S}
\tabularnewline
\cline{3-8}
	  & & & Total & & Total & &  
\tabularnewline
\hline
\hline
\multicolumn{8}{|c|}{General case (splitting in two parts) \footnote{table extracted from \cite{koseleff1993}}} \tabularnewline
\hline

1 	& 2     & 2     & 2     & 2		&    	& 2			&    \tabularnewline
2 	& 1     & 1 	& 3     &   	& 2  	& 3			& 3   \tabularnewline
3 	& 2     & 2 	& 5     & 2  	&    	& 6			&    \tabularnewline
4 	& 3     & 3 	& 8     &   	& 4  	& 7			& 7   \tabularnewline
5 	& 6   	& 6 	& 14    & 6 	&    	& 			&    \tabularnewline
6 	& 9   	& 9 	& 23    &  		& 10 	& 			& 15   \tabularnewline
7 	& 18   	& 18 	& 41    & 18  	&    	& 			&    \tabularnewline
8 	& 30   	& 30 	& 71	&   	& 28 	& 			& 31   \tabularnewline
9 	& 56   	& 56 	& 127 	& 56 	&    	& 			&    \tabularnewline
10 	& 99   	& 99 	& 226  	&  		& 84 	& 			&    \tabularnewline
\hline
\hline
\multicolumn{8}{|c|}{General case (splitting in three parts)} \tabularnewline
\hline
1 	& 3     	& 3     & 3     & 3		&    	& 3			&    \tabularnewline
2 	& 3     	& 3 	& 6     &   	& 3  	& 5			& 5   \tabularnewline
3 	& 8     	& 8 	& 14    & 8  	&    	& 			&    \tabularnewline
4 	& 18     	& 18 	& 32    &   	& 11  	& 			& \footnote{\citep{koseleff1993,koseleff1996,tang2002}}13   \tabularnewline
5 	& 48  		& 48 	& 80    & 48 	&    	& 			&    \tabularnewline
6 	& 116   	& 116 	& 196   &  		& 59 	& 			& \footnote{\citep{yoshida1990}}29   \tabularnewline
7 	& 312   	& 312 	& 508   & 312  	&    	& 			&    \tabularnewline
8 	& 810   	& 810 	& 1318	&   	& 371 	& 			& \footnote{\citep{yoshida1990}}61    \tabularnewline
9 	& 2184   	& 2184 	& 3502 	& 2184 	&    	& 			&    \tabularnewline
10 	& 5880   	& 5880 	& 9382  &  		& 2555 	& 			& \footnote{\citep{sofroniou2005}}125   \tabularnewline
\hline
\hline
\multicolumn{8}{|c|}{Free rigid body (splitting in two parts, RS)} \tabularnewline
\hline
1 	& 3 	& 2 	& 2 	& 2		&   	& 2	 &    \tabularnewline
2 	& 1 	& 1 	& 3 	&  		& 2 	& 3  & 3  \tabularnewline
3 	& 2 	& 2 	& 5 	& 2		&   	& 5  &    \tabularnewline
4 	& 3 	& 3 	& 8 	&  		& 4 	& 8  & 7  \tabularnewline
5 	& 6 	& 6 	& 14 	& 6		&   	& 14 &    \tabularnewline
6 	&  	& 6 	& 20 	&  		& 10 	& 20 & 19 \tabularnewline
7 	& 	& 11	& 31	& 11	& 		& 31 &    \tabularnewline
8 	& 	& 10	& 41	&		& 21	& 41 & 41 \tabularnewline
9 	& 	& 17	& 58	& 17	& 		& 58 &    \tabularnewline
10 	& 	& 15	& 73	&		& 38	& 73 & 75 \tabularnewline
\hline
\hline
\multicolumn{8}{|c|}{Free rigid body (splitting in three parts, ABC)} \tabularnewline
\hline
1 	& 3 	& 3 	& 3 	& 3  	&   	& 3	 &    \tabularnewline
2 	& 1 	& 1 	& 4 	&  		& 3 	& 4  & 5  \tabularnewline
3 	& 2 	& 2 	& 6 	& 2  	&   	& 6  &    \tabularnewline
4 	& 3 	& 3 	& 9 	&  		& 5 	& 9  & 9  \tabularnewline
5 	& 6 	& 6 	& 15 	& 6  	&   	& 15 &    \tabularnewline
6 	&  	& 6 	& 21 	&  		& 11 	& 21 & 21 \tabularnewline
7 	& 	& 11	& 32	& 11	& 		& 32 &    \tabularnewline
8 	& 	& 10	& 42	&		& 22	& 42 & 43 \tabularnewline
9 	& 	& 17	& 59	& 17	& 		& 59 &    \tabularnewline
10 	& 	& 15	& 74	&		& 39	& 74 & 77 \tabularnewline
\hline
\end{tabular}
\caption{\label{table:conditions}Dimension of Lie algebras $\mathcal{L}_k$, number of equations, number of stages of an integrator non symmetric (NS) and symmetric (S) in the case of a general Hamiltonian and of the one of the free rigid body with respect to the order for a decomposition in two and three parts.}
\end{minipage}
\end{table}  

For the even orders, the number of monomials of the linear combination of the theorem 1 is $n_{2k}=k(k+1)/2$ for $k\geq1$.
For the odd orders, the number of monomials of the linear combination of the theorem 1 is $n_{2k+1}=k(k+5)/2$ for $k\geq1$.
For the first order, the number of monomials is three.
To obtain the number of equations to verify to build an integrator, we sum the number of monomials until the order of the integrator.
Moreover from the theorem 2, we have one additional relation between the coefficients of each odd order. For the splitting RS, we have the additional relation $\b_{1,010}=\b_{1,001}$. Each relation reduces by one the number of equations to verify in order to cancel the remainder of these orders.

We obtain the number of equations to verify for an integrator of order $2n$
\begin{equation}
N\left(2n\right)  =  \frac{n\left(2n^2+9n-11\right)}{6}+4+(-1)_{RS}\label{eq:Ntoteven}
\end{equation}
and the number of equations for an integrator of order $2n+1$
\begin{equation}
N\left(2n+1\right) = \frac{n\left(n^2+6n+2\right)}{3}+3+(-1)_{RS}.\label{eq:Ntotodd}
\end{equation}
$(-1)_{RS}$ indicates that for the splitting RS the number of equations must be reduced by $1$.
The number of equations for a symmetric integrator of order $2n$ is
\begin{equation}
N_{sym}\left(2n\right)=\frac{n\left(n^2+6n-13\right)}{6}+4+(-1)_{RS}.\label{eq:Ntotsym}
\end{equation}

Table \ref{table:conditions}, whose the part on the ordinary Hamiltonian split in two parts is extracted from \cite{koseleff1993}, compares for the two splittings the dimensions of the Lie algebras $\mathcal{L}_k$ for the Hamiltonian of the free rigid body and for an ordinary Hamiltonian and precises the number of equations and the number of stages.
Eqs. (\ref{eq:Ntoteven}) and (\ref{eq:Ntotodd}) allow us to fill in table \ref{table:conditions} the column 7 and Eq. (\ref{eq:Ntotsym}) the column 8 for the free rigid body.
We note that for the five first orders these formulas give results which correspond to the minimal number of stages given by the dimensions of the Lie algebras $\mathcal{L}_k$.
Therefore, with these formulas we obtain the minimal number of stages for the orders $1$, $2$, $3$, $4$ and $5$.
The dimensions of the Lie algebras in the general case have been obtained by \cite{koseleff1993}, \cite{mclachlan1995} and \cite{koseleff1996} for a Hamiltonian split in two parts and by \cite{munthe-kaas1999} for a Hamiltonian split in three parts.
To fill the table \ref{table:conditions}, we also use the inventory of the splitting integrators made by \cite{blanes2008} and \cite{skokos2014} for Hamiltonians which can be split in respectively two and three parts.

For an ordinary Hamiltonian split in two parts, \cite{koseleff1996} demonstrated that the minimal number of stages for a symmetric 4th-order integrator is 7, where the Yoshida's scheme is the only real solution. In the special case of the free rigid body, the number of stages is 7 (table \ref{table:conditions}) and the splitting RS cannot benefit from the algebra of the angular momentum to construct integrators with fewer stages. For an ordinary Hamiltonian split in three parts, \cite{koseleff1996} demonstrated that the minimal number of stages for a symmetric 4th-order integrator is 13 and \cite{tang2002} proved that the Yoshida's scheme is the only real solution. For the free rigid body, the number of stages becomes 9 (table \ref{table:conditions}) and the splitting ABC profits by the algebra of the free rigid body to construct integrators with fewer stages.

\section{Construction of symmetric integrators\label{sec:construction}}

In this part, we explain how to construct symplectic integrators for the rigid body which benefit from the algebra of the angular momentum.
We limit ourselves to 4th-order schemes, which are the easiest to construct.   

\subsection{Splitting RS\label{sec:construction_RS}}

For a Hamiltonian split into two parts, we have seen in section \ref{sec:constraint_number} that the minimum number of stages for a 4th-order symmetric integrator of the free rigid body is 7.
Two splitting schemes of 7 stages exist
\begin{equation}
\begin{array}{ll}
RSRSRSR & \\
SRSRSRS & \\
\end{array}
\end{equation}
where the Hamiltonians $R$ and $S$ are defined respectively in Eqs. (\ref{eq:step_R}) and (\ref{eq:step_S}). In order to compute the coefficients of these two integrators, we scale everything by $I_1$ and write the Hamiltonian of the free rigid body as
\begin{equation}
H= \frac{G_1^2}{2I_1}+  \frac{G_2^2}{2I_2} + \frac{G_3^2}{2I_3}=\frac{1}{2I_1}\left(G_{1}^{2}\left(1-x\right)+G_{3}^{2}\left(y-x\right)+xG^{2}\right),
\end{equation}   
with
\begin{equation}
 x = \frac{I_1}{I_2}, \ \qquad   y = \frac{I_1}{I_3}.
\end{equation}

\subsubsection{Computation of the coefficients}

We start with the scheme RSRSRSR defined by
\begin{equation}
\mathcal{S}_{RSRSRSR}\left(h\right)=e^{a_{1}hL_{R}}e^{b_{1}hL_{S}}e^{a_{2}hL_{R}}e^{b_{2}hL_{S}}e^{a_{2}hL_{R}}e^{b_{1}hL_{S}}e^{a_{1}hL_{R}}.
\end{equation}
The 1st-order conditions impose
\begin{equation}
 2a_1+2a_2=1, \  \qquad   2b_1+b_2=1. \ 
\end{equation}
We have only two free parameters $a_1$ and $b_1$, and the other coefficients are given by
\begin{equation}
a_2=\frac{1}{2}-a_1, \  \qquad   b_2=1-2b_1. \ 
\end{equation}
With the Baker-Campbell-Hausdorff formula, we can compute the Hamiltonian $K$ which is effectively integrated for the integrator $\mathcal{S}_{RSRSRSR}\left(h\right)=e^{L_K}$ with
\begin{equation}
hK=hH+\frac{h^3}{\left(2I_1\right)^3}\left(P'_1G_1^2G_2^2 + P'_2G_1^2G_3^2 + P'_3G_2^2G_3^2\right)+O\left(h^5\right),
\end{equation}
and where the coefficients $P'_i=P_i(2I_1)^3$ are given by
\begin{equation}
\EQM{
P'_1 =& \frac{1}{3}(x-1)^2(x-y)\left(1-6(1-2 a_1)^2 b_1\right) \crm
P'_2 =& -\frac{1}{3}(x-1)(x-y)\left((x-1)(1-6(1-2 a_1)^2b_1)\right.\crm
&\left.+2(x-y)\left(6\left(1-2 a_1\right)\left(1-b_1\right)b_1-1\right)\right)\crm
P'_3 =& \frac{2}{3}(x-1)(x-y)^2 \left(6(1-2 a_1)(1-b_1) b_1-1\right). \crm 
}
\end{equation}
With $P'_1+P'_2+P'_3=0$, it is sufficient to impose $P'_1=P'_3=0$ to cancel the 3rd-order remainder. For an asymmetric body ($x\neq 1$, $x\neq y$, $y\neq 1$), we obtain  
\begin{equation}
b_1^3-2b_1^2 + b_1-\frac{1}{6} =0 , \  \qquad   a_1 = \frac{1}{2}b_1. \ 
\end{equation}
The only real solution is the Yoshida's scheme with $b_1=1/(2-2^{1/3})$ \citep{yoshida1990}.

With the scheme SRSRSRS defined by
\begin{equation}
\mathcal{S}_{SRSRSRS}\left(h\right)=e^{a_{1}hL_{S}}e^{b_{1}hL_{R}}e^{a_{2}hL_{S}}e^{b_{2}hL_{R}}e^{a_{2}hL_{S}}e^{b_{1}hL_{R}}e^{a_{1}hL_{S}},
\end{equation}
we have
\begin{equation}
\EQM{
P'_1=& \frac{2}{3}(x-1)^2 (x-y) \left(6(1-2 a_1)(1-b_1)b_1-1\right)\crm
P'_2=&-\frac{1}{3}(x-1)(x-y)\left(2(x-1)(6 (1-2 a_1)(1-b_1)b_1-1)\right.\crm
&\left.+(x-y)(1-6 (1-2 a_1)^2b_1)\right) \crm
P'_3=& \frac{1}{3}(x-1)(x-y)^2 \left(1-6 (1-2 a_1)^2 b_1\right). \crm 
}
\end{equation}
Like previously, there is only one real solution, which is the Yoshida's one.

Therefore, using the algebra of the angular momentum for the splitting RS does not allow us to obtain supplementary integrators other than Yoshida's scheme for the integrators with seven stages.

\subsubsection{Notes on the splitting RS\label{sec:construction_RS_note}}

We observe that if we switch the parts $G_1^{2}/(2I_1)$ and $G_3^{2}/(2I_3)$ for the scheme $RSRSRSR$, we obtain the same 3rd-order remainder as the scheme $SRSRSRS$.
This is due to the fact that $R_G=G^{2}/(2I_2)$ commutes with $R_1=G_1^2(1/(2I_1)-1/(2I_2))$ and $R_3=G_3^2(1/(2I_3)-1/(2I_2))$ (This commutation has been previously noted by \cite{fasso2003}.).
Therefore the scheme $RSRSRSR$ with the permutation of the parts $G_1^{2}/(2I_1)$ and $G_3^{2}/(2I_3)$ has a remainder identical to the one of the scheme $SRSRSRS$ for any order.

\cite{fasso2003} has already noted for the leapfrog scheme that the 3rd-order remainder of the type $RS$ with this permutation is identical to the one of the $SR$.
Here we see that this can be generalized to any order and any scheme.
Therefore it is not necessary to consider the schemes $SR$, which in the case of a kinetic energy coupled to a potential part is more expensive than the schemes $RS$, because the stage $R$ is cheaper than $S$.

As the part $R_G$ commutes with the parts $R_1$ and $R_3$, it is possible to gather all the stages of type $e^{L_{R_G}}$ in a step of integration to decrease the computation time.
An integrator with the splitting RS has then only a rotation around the angular momentum by step of integration.
This reduction of the computation time for the splitting RS was not previously noticed as far as we know.

\subsection{Splitting ABC: integrator N\label{sec:construction_N}}

For a Hamiltonian split into three parts, we have seen in section \ref{sec:constraint_number} that the minimum number of stages for a 4th-order symmetric integrator of the free rigid body is 9.
Seven splitting schemes of 9 stages and beginning with the stage $A$ and followed by the stage $B$, exist. We call them integrators N and they can be sorted in alphabetical order as
\begin{equation}
\begin{array}{ll}
ABABCBABA & N1  \\
ABACACABA & N2  \\
ABACBCABA & N3  \\
ABCABACBA & N4  \\
ABCACACBA & N5  \\
ABCBABCBA & N6  \\
ABCBCBCBA & N7.  \\ 
\end{array} \label{eq:schemes}
\end{equation}
To obtain all the possible schemes, we consider the six permutations $ABC$, $BCA$, $CAB$, $ACB$, $CBA$, $BAC$ as \cite{fasso2003}.
For example, the scheme N4 with the permutation $CAB$ becomes $CABCACBAC$. Therefore we count 42 4th-order possible schemes.

To compute the coefficients of each integrator, we scale everything by $I_1$ and write 
\begin{equation}
H= \frac{G_1^2}{2I_1}+  \frac{G_2^2}{2I_2} + \frac{G_3^2}{2I_3}=\frac{1}{2I_1} \left(G_1^2+\left(1+x\right)G_2^2+\left(1+y\right)G_3^2\right),
\end{equation}
with
\begin{equation}
 1+x = \frac{I_1}{I_2}, \  \qquad  1+y = \frac{I_1}{I_3}. \ 
\label{xythree}
\end{equation}

\subsubsection{Computation of the coefficients}

We explain here in detail the computation of the coefficients only for the scheme N4 with the permutation $ABC$
\begin{equation}
\mathcal{S}_{N4\,ABC}\left(h\right)=e^{a_{1}hL_{A}}e^{b_{1}hL_{B}}e^{c_{1}hL_{C}}e^{a_{2}hL_{A}}e^{b_{2}hL_{B}}e^{a_{2}hL_{A}}e^{c_{1}hL_{C}}e^{b_{1}hL_{B}}e^{a_{1}hL_{A}}.
\end{equation}
The conditions of the 1st-order impose
\begin{equation}
2a_1+2a_2=1, \  \qquad 2b_1+b_2=1, \  \qquad   2c_1=1. \ 
\end{equation}
We have only two free parameters $a_1$ and $b_1$ and the other coefficients are given by 
\begin{equation}
a_2=\frac{1}{2}-a_1, \  \qquad b_2=1-2b_1, \  \qquad   c_1=\frac{1}{2}. \ 
\end{equation}
With the Baker-Campbell-Hausdorff formula, we determine the Hamiltonian $K$ effectively integrated of the scheme $\mathcal{S}_{N4\,ABC}\left(h\right)=e^{hL_K}$ with
\begin{equation}
hK=hH+\frac{h^3}{\left(2I_1\right)^3}\left(P'_1 G_1^2G_2^2 + P'_2 G_1^2G_3^2 + P'_3 G_2^2G_3^2\right)+O\left(h^5\right),
\end{equation}
where $P'_1$, $P'_2$ and $P'_3$ depend on $a_1$, $b_1$, $x$ and $y$ with
\begin{eqnarray}
\EQM{
P'_1&=&-\frac{1}{3}+2b_1-8a_1b_1+8a_1b_1^{2}+4a_1^{2}-8a_1^{2}b_1+4yb_1^{2}-8ya_1b_1+4ya_1^{2}-x+2xb_1 \crm
&&+4xa_1-16xa_1b_1+16xa_1b_1^{2}-8xa_1^{2}b_1-4xyb_1+8xyb_1^{2}+4xya_1-8xya_1b_1\crm
&&-8x^{2}a_1b_1+8x^{2}a_1b_1^{2}+\frac{2}{3}x^{2}y-4x^{2}yb_1+4x^{2}yb_1^{2},\crm
P'_2&=&-\frac{1}{3}+2a_1-4a_1^{2}+8a_1^{2}b_1+8ya_1b_1-4ya_1^{2}+2y^{2}b_1\crm
&&-2y^{2}a_1+8xa_1^{2}b_1+8xya_1b_1-\frac{1}{3}xy^{2}+2xy^{2}b_1,\crm
P'_3&=& \frac{2}{3}-2b_1-2a_1+8a_1b_1-8a_1b_1^{2}-4yb_1^{2}-2y^{2}b_1+2y^{2}a_1+x-2xb_1\crm
&&-4xa_1+16xa_1b_1-16xa_1b_1^{2}+4xyb_1-8xyb_1^{2}-4xya_1+\frac{1}{3}xy^{2}-2xy^{2}b_1\crm
&&+8x^{2}a_1b_1-8x^{2}a_1b_1^{2}-\frac{2}{3}x^{2}y+4x^{2}yb_1-4x^{2}yb_1^{2}.\crm
}\label{eq:exp_Pk}
\end{eqnarray}
With $P'_1+P'_2+P'_3=0$, we only need to cancel the coefficients $P'_1$ and $P'_2.$ Using Gr\"{o}bner base reduction \citep{buchberger1965}, solving $P'_k=0$ for $k=1,2$ can be reduced to the two equations 
\begin{equation}
\EQM{
\a a_1^3 + \b a_1^2 + \c a_1 + \d  =0 \crm
\a' a_1 + \b' b_1 + \c' = 0,
\label{eqcoef}
}
\end{equation}
where $\a$, $\b$, $\c$, $\d$, $\a'$, $\b'$, $\c'$ are constant coefficients that depend on the moments of inertia with
\begin{equation}
\EQM{
\a&=   48\,{x}^{2}y-24\,x{y}^{2}+144\,xy-72\,{y}^{2}+72\,x+24,\crm
\b&=   48\,{x}^{2}{y}^{2}-12\,x{y}^{3}+144\,x{y}^{2}-36\,{y}^{3}+36\,xy+72\,{y}^{2}-24\,x+60\,y,\crm
\c&=   12\,{x}^{2}{y}^{3}+24\,x{y}^{3}-24\,x{y}^{2}+18\,{y}^{3}-24\,xy-18\,{y}^{2}-30\,y-6,\crm
\d &= {x}^{2}{y}^{4}-6\,x{y}^{3}-4\,x{y}^{2}-3\,{y}^{3}+3\,y+1, \crm
\a'&=  4\,{x}^{2}y-2\,x{y}^{2}+12\,xy-6\,{y}^{2}+6\,x+2,\crm
\b'&=  2\,{x}^{2}{y}^{2}+8\,x{y}^{2}+6\,xy+6\,{y}^{2}+2\,x+6\,y+2,\crm
\c'&= -x{y}^{2}-3\,xy-2\,x-y-1. 
}
\end{equation}
These two equations can count several solutions, which depend on the moments of inertia through $x$ and $y$.
The coefficients of these integrators are specific to a sorted triplet of moments of inertia.
For each different triplet of moments of inertia, we need to compute the coefficients of the integrator. In exchange for this dependence of coefficients, we have integrators with fewer stages.

The coefficients $P'_k$ of the equations (\ref{eq:exp_Pk}) can be used for any permutation of the moments of inertia of the integrator N4.
Instead of doing a permutation of the stages $A$, $B$, $C$, we can switch the moments of inertia and modify the values of $x$ and $y$.
In the appendix \ref{AppendixA}, we indicate the equations to solve to obtain the coefficients of the seven integrators N.

\subsubsection{Decreasing of the number of stages for specific bodies}

For the integrator N4, there are values of ($x$,$y$) for whom $\d=0$ and then $a_1=0$ can be a solution.
There is then a curve ($\d=0$) in the set of definition of $(x,y)$ where it is possible to obtain 4th-order integrators with 7 or even 5 stages for some discrete values of this curve where $\c'=0$ and where $a_1=b_1=0$ is a solution.
\cite{fasso2003} observed a similar result and found that the leapfrog scheme becomes a 4th-order integrator for the case of a flat body with moments of inertia $\left(0.25,0.75,1\right)$.  

\subsection{Estimation of the remainder \label{sec:construction_analytical}}

We have 42 possible integrators of type N, and for each of them, several solutions for their coefficients.
To determine the best integrator, we can perform numerical integrations.
Alternatively, we can also estimate faster the precision of the integrator by evaluating the analytical remainder.
For an integrator of order $n$, the precision can be estimated by the Euclidean norm of the remainder terms of lowest degree $n+1$.

From theorem 1, the 5th-order remainder can be written
\begin{equation}
H_{R_{5}}=Q_{1}G_{1}^{2}G_{2}^{4}+Q_{2}G_{2}^{2}G_{3}^{4}+Q_{3}G_{3}^{2}G_{1}^{4}+Q_{4}G_{1}^{4}G_{2}^{2}+Q_{5}G_{2}^{4}G_{3}^{2}+Q_{6}G_{3}^{4}G_{1}^{2}+Q_{7}G_{1}^{2}G_{2}^{2}G_{3}^{2},\label{eq:H5_2}
\end{equation} 
and from theorem 2, we verify the reduction formula for the 5th-order
\begin{equation}
3\left(Q_{1}+Q_{2}+Q_{3}+Q_{4}+Q_{5}+Q_{6}\right)+Q_{7}=0. \label{eq:Q_k=0}
\end{equation} 
$H_{R_5}$ belongs then to the set $\mathcal{V}$ of basis ($G_{1}^{2}G_{2}^{4}$, $G_{2}^{2}G_{3}^{4}$, $G_{3}^{2}G_{1}^{4}$, $G_{1}^{4}G_{2}^{2}$, $G_{2}^{4}G_{3}^{2}$, $G_{3}^{4}G_{1}^{2}$, $G_{1}^{2}G_{2}^{2}G_{3}^{2}$).
The estimation of the remainder for a 4th-order integrator by the Euclidean norm $\norm{H_{R_{5}}}$ in the set $\mathcal{V}$ is then
\begin{equation}
\norm{H_{R_{5}}}=\sqrt{Q_{1}^{2}+Q_{2}^{2}+Q_{3}^{2}+Q_{4}^{2}+Q_{5}^{2}+Q_{6}^{2}+Q_{7}^{2}}. \label{eq:res5}
\end{equation} 

\subsection{Additional parameter\label{sec:construction_additional}}

In order to lower the value of the remainder, it is possible to add to any of the previous schemes an additional stage (two stages for a symmetric integrator). 
This provides an additional free parameter that we can determine in order to minimize the remainder of Eq. (\ref{eq:res5}). 

\subsubsection{Splitting RS: integrator R}

The addition of a parameter in the previous 4th-order RS example gives the integrator of nine stages
\begin{eqnarray}
RSRSRSRSR \nonumber \\
\end{eqnarray}
which we call integrator R.
The scheme $\mathcal{S}_R(h)=e^{hL_K}$ is given by
\begin{equation}
e^{a_{1}hL_{R}}e^{b_{1}hL_{S}}e^{a_{2}hL_{R}}e^{b_{2}hL_{S}}e^{a_{3}hL_{R}}e^{b_{2}hL_{S}}e^{a_{2}hL_{R}}e^{b_{1}hL_{S}}e^{a_{1}hL_{R}}
\end{equation}
with 
\begin{equation}
a_3=1-2\left(a_1+a_2\right),\ \qquad b_2=\frac{1}{2}-b_1,\
\end{equation}
and
\begin{equation}
hK=hH+h^3H_{R_3}+h^5H_{R_5}+O\left(h^7\right).\
\end{equation}
With $x=I_1/I_2$ and $y=I_1/I_3$, we solve $P_1=P_2=0$ to cancel $H_{R_3}$ and minimize $\norm{H_{R_5}}^2$ using Lagrange multipliers.

\subsubsection{Splitting ABC: integrator P} 

For the splitting ABC, we have the fifteen following possible schemes
\begin{equation}
\begin{array}{ll}
ABABACABABA & P1  \\
ABABCACBABA & P2  \\
ABABCBCBABA & P3  \\
ABACABACABA & P4  \\
ABACACACABA & P5  \\
ABACBABCABA & P6  \\
ABACBCBCABA & P7  \\ 
ABCABABACBA & P8  \\
ABCABCBACBA & P9  \\
ABCACACACBA & P10 \\
ABCACBCACBA & P11 \\
ABCBABABCBA & P12 \\
ABCBACABCBA & P13 \\
ABCBCACBCBA & P14 \\
ABCBCBCBCBA & P15 \\ 
\end{array} 
\end{equation}
which we call integrators P. The first scheme $\mathcal{S}_{P1\,ABC}\left(h\right)=e^{hL_K}$ is given by
\begin{equation}
e^{a_{1}hL_{A}}e^{b_{1}hL_{B}}e^{a_{2}hL_{A}}e^{b_{2}hL_{B}}e^{a_{3}hL_{A}}e^{c_{1}hL_{C}}e^{a_{3}hL_{A}}e^{b_{2}hL_{B}}e^{a_{2}hL_{A}}e^{b_{1}hL_{B}}e^{a_{1}hL_{A}}
\end{equation}
with 
\begin{equation}
a_3=\frac{1}{2}-\left(a_1+a_2\right),\ \qquad b_2=\frac{1}{2}-b_1, \ \qquad c_1=1,
\end{equation}
and
\begin{equation}
hK=hH+h^3H_{R_3}+h^5H_{R_5}+O\left(h^7\right).
\end{equation}
With $1+x=I_1/I_2$ and $1+y=I_1/I_3$, we solve $P_1=P_2=0$ to cancel $H_{R_3}$ and minimize $\norm{H_{R_5}}^2$.

\section{Numerical tests\label{sec:numerical}}

In this part, we realize numerical tests to compare the efficiency of the obtained dedicated integrators to the one of the usual symplectic integrators.

\subsection{Method\label{sec:numerical_method}}

To compare the integrators, we need to evaluate their cost $\mathcal{C}$ for the same step size $h$.  
We can then compare two integrators of the same accuracy by comparing their reduced step size defined as in \cite{farres2013} by $S(h)={h}/{\mathcal{C}}$. 
We note $\mathcal{C}_0$ the numerical cost of a rotation.
The stages $e^{hL_{A}}$, $e^{hL_{B}}$, $e^{hL_{C}}$, $e^{hL_{R}}$ correspond to one rotation around a principal axis and have the same cost $\mathcal{C}_0$.
Therefore a scheme ABC with $N$ stages of type $e^{hL_A}$, $e^{hL_{B}}$, $e^{hL_{C}}$ has a cost of $N\mathcal{C}_0$.
However the stage $e^{hL_{S}}$ is composed of one rotation around one of the principal axes and one rotation around the angular momentum. 
The cost of $e^{hL_{S}}$ is thus $2\mathcal{C}_0$.
With the reduction of the computation time noted in section \ref{sec:construction_RS_note} for the splitting RS, we need to perform the rotation around the angular momentum only one time by integration step.
Therefore a scheme RS with $N$ stages of type $e^{hL_{R}}$, $e^{hL_{S}}$ has a cost of $(N+1)\mathcal{C}_0$.

In all the numerical tests, we start with a body of angular momentum $\mathbf{G}=\left(1,1,1\right)$ for an initial orientation of the body $(\mathbf{I},\mathbf{J},\mathbf{K})=(\left(1,0,0\right),\left(0,1,0\right),\left(0,0,1\right))\in\mathbb{R}^9$.
We perform numerical integrations over a period $T=1$ with a step size $h_{i}={1}/{2^{i}}$ for $i=1,\ldots,10$.

For all the numerical tests, we estimate the precision of an integrator by the numerical remainder $R_n$ on the orientation of the body, which is obtained by the following procedure.
At every step, we compute the Euclidean norm in $\mathbb{R}^9$ of the difference between the orientation of the body computed with the given symplectic scheme and the one given by the exact solution.
To compute the exact solution of the free rigid body, we use the matrix algorithm of \cite{celledoni2008}. 
For small angles, this difference can be interpreted as the quadratic mean of the angular errors on the three principal axes multiplied by $\sqrt{3}$.
The numerical remainder $R_n$ is then obtained by taking the average of this difference on the whole integration.    

We compare the efficiency of the schemes developed for the rigid body to the general integrators which can be used for any Hamiltonian.
Among them, \cite{mclachlan1995} distinguished the symmetric integrators (type S) and the symmetric integrators built from symmetric integrators of lower order (type SS).
We have seen in section \ref{sec:construction_RS_note} that it is not necessary to consider the schemes $SR$ provided that we consider all the possible permutations of the moments of inertia for the schemes $RS$.

The reference schemes used for the tests are the classical leapfrog scheme (ABCBA2, RSR2), the Yoshida's 4th-order scheme with a composition of three leapfrog schemes (ABCBA4 SS3 Yoshida, RSR4 SS3 Yoshida) \citep{yoshida1990}, the Suzuki's 4th-order scheme  with a composition of five leapfrog schemes (ABCBA4 SS5 Suzuki, RSR4 SS5 Suzuki) \citep{suzuki1990}, the McLachlan's 4th-order scheme with a composition of five leapfrog schemes (ABCBA4 SS5 McLachlan, RSR4 SS5 McLachlan) and the McLachlan's 4th-order symmetric schemes of $2n+1$ stages with $n=4$ (ABC4 S4 McLachlan, RS4 S4 McLachlan) and $n=5$ (ABC4 S5 McLachlan, RS4 S5 McLachlan) \citep{mclachlan1995}.

\begin{figure}
\centering
\input{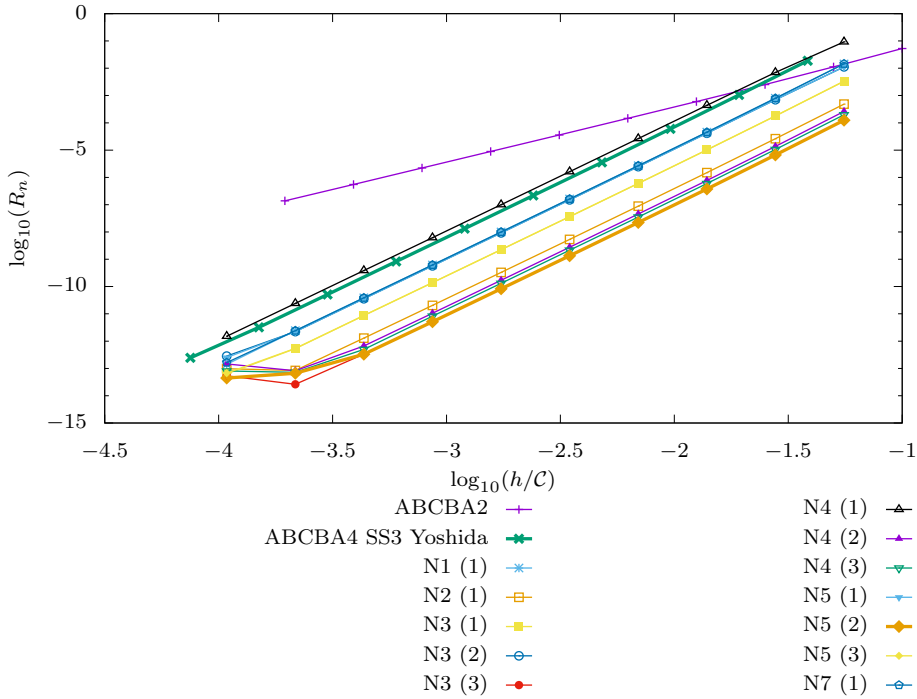} 
\caption{\label{fig:res_sphere} Comparison of the numerical remainders $R_n$ for a spherical top between the integrators of the table \ref{table:sol_sphere}, the classical leapfrog scheme (ABCBA 2) and the Yoshida's scheme (ABCBA4 SS3 Yoshida).}
\end{figure}

\begin{figure}
\centering
\input{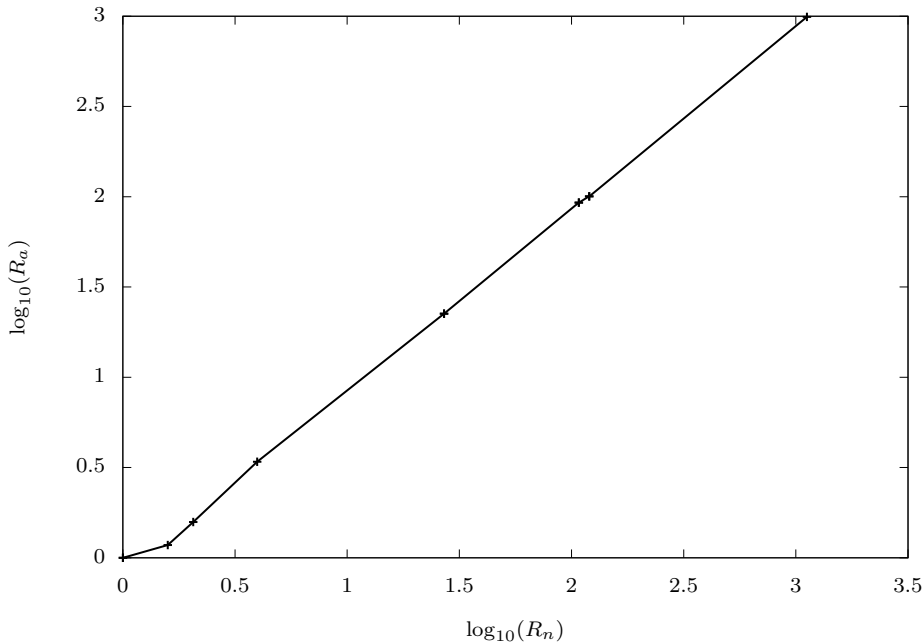} 
\caption{\label{fig:res_sphere2} Normalized analytical remainder $R_a$ (table 3) with respect to the normalized numerical remainder $R_n$ (table 3) for the integrators of the table \ref{table:sol_sphere}.}
\end{figure}

\subsection{Spherical top\label{sec:numerical_sphere}}

\begin{table}
\centering
\begin{tabular}{|c|l|}
\hline
Integrator & Coefficients \\
\hline
N1 (1) & $a_1=-8.5120719195965763404768780897146082692210\times10^{-1}$ \\ 
 & $b_1=-1.7560359597982881702384390448573041346105\times10^{-1}$ \\
 \hline
N2 (1) & $a_1=\frac{1}{6}$ \\ 
 & $a_2=\frac{1}{6}$ \\
 \hline
N3 (1) & $a_1=\frac{1}{\sqrt{3}} \cos \frac{17\pi}{18}$ \\ 
 & $b_1=\frac{3}{2}-a_1-6a_1^{2}$ \\
N3 (2) & $a_1=\frac{1}{\sqrt{3}} \cos \frac{7\pi}{18}$ \\ 
 & $b_1=\frac{3}{2}-a_1-6a_1^{2}$ \\
N3 (3) & $a_1=\frac{1}{\sqrt{3}} \cos \frac{5\pi}{18}$ \\ 
 & $b_1=\frac{3}{2}-a_1-6a_1^{2}$ \\
 \hline
N4 (1) & $a_1=\frac{1}{\sqrt{3}} \cos \frac{17\pi}{18}$ \\ 
& $b_1=\frac{1}{2}-a_1$ \\
N4 (2) & $a_1=\frac{1}{\sqrt{3}} \cos \frac{7\pi}{18}$ \\ 
& $b_1=\frac{1}{2}-a_1$ \\
N4 (3) & $a_1=\frac{1}{\sqrt{3}} \cos \frac{5\pi}{18}$ \\ 
& $b_1=\frac{1}{2}-a_1$ \\
 \hline
N5 (1) & $a_1=\frac{1}{\sqrt{3}} \cos \frac{17\pi}{18}$ \\ 
& $c_1=-\frac{1}{2}+2a_1+6a_1^{2}$ \\
N5 (2) & $a_1=\frac{1}{\sqrt{3}} \cos \frac{7\pi}{18}$ \\ 
& $c_1=-\frac{1}{2}+2a_1+6a_1^{2}$ \\
N5 (3) & $a_1=\frac{1}{\sqrt{3}} \cos \frac{5\pi}{18}$ \\ 
& $c_1=-\frac{1}{2}+2a_1+6a_1^{2}$ \\
 \hline
N7 (1) & $b_1=6.7560359597982881702384390448573041346096\times10^{-1}$ \\
& $c_1=1.3512071919596576340476878089714608269219$ \\
\hline
\end{tabular}
\caption{\label{table:sol_sphere} Coefficients of the 4th-order integrators N for the spherical top.}
\end{table}    
\begin{table}
\centering
\begin{tabular}{rrr}
\hline
Integrator & Numerical remainder & Analytical remainder \\
\hline
N1 (1) & 119.85 	& 100.50 \\
N2 (1) & 3.97	& 3.40 \\
N3 (1) & 27.04 	& 22.51 \\
N3 (2) & 107.78 	& 92.79 \\
N3 (3) & 1.00 	& 1 \\
N4 (1) & 1120.50 & 991.42 \\
N4 (2) & 2.06 	& 1.58\\
N4 (3) & 1.59 	& 1.18 \\
N5 (1) & 107.82 	& 92.79 \\
N5 (2) & - 			& - \\
N5 (3) & 27.02 	& 22.51 \\
N7 (1) & 120.07 	& 100.50 \\
\hline
\end{tabular}
\caption{\label{table:res_sphere}Numerical and analytical remainders for the integrators N dedicated to the spherical top normalized by those of the best integrator N5 (2).}
\end{table}

The moments of inertia of a spherical top are identical ($I_1=I_2=I_3=1$).
The rotation of a spherical top is then trivially integrable and the exact solution corresponds to a rotation around the angular momentum. However this simple example allows us to test our algorithms and to find an efficient way to compare the different integrators. 
In Eq. (\ref{xythree}), we have $x=y=0$ and the Hamiltonian reduces to $H= (G_1^2+G_2^2+G_3^2)/2$.

The RS schemes are trivial as $R$ reduces to the identity, and $S$ to the rotation around the angular momentum.
We will thus only consider here the splitting ABC.
For the integrators N of Eq. (\ref{eq:schemes}), we obtain 12 real solutions (table \ref{table:sol_sphere}).
As all the moments of inertia are identical, it is not necessary to consider their permutations.
To compare these  solutions, the numerical remainder (section \ref{sec:numerical_method}) is compared to the reduced step $S(h)=h/\mathcal{C}$ for all the integrators N of table \ref{table:sol_sphere} on figure \ref{fig:res_sphere}.
We sort the solutions in ascending order of the coefficient $a_1$.
The same quantities are plotted for the classical leapfrog scheme of order 2, and the Yoshida's integrator of order 4 \citep{yoshida1990}.
The best integrator for the spherical top is N5 (2) which is, at equivalent reduced cost $h/\mathcal{C}$, about 700 times more accurate than the Yoshida's integrator.

On table \ref{table:res_sphere}, the estimation of the analytical remainder of section \ref{sec:construction_analytical} is also provided in column 3 and they are very well correlated to the numerical remainders (figure \ref{fig:res_sphere2}). In the following, we can then use the analytical remainder as a fast estimate of the accuracy of the integrators. 

\subsection{Water molecule}

\begin{table}
\centering
\begin{tabular}{|c|l|p{1.35cm}|}
\hline
Integrator & Coefficients & Analytical remainder ($\times10^{-2}$) \\
\hline
N1 ABC (1) & $a_1=2.3009531403182120088035835866488791305526\times10^{-1}$ & 2.09\\ 
    & $b_1=2.7028961116588991802338916444296474989967\times10^{-1}$ & \\
N1 ABC (2) & $a_1=3.1275929803539412927331110867593240725096\times10^{-1}$ &  3.83\\ 
   & $b_1=1.8915198437863547818626148139249232203935\times10^{-1}$ & \\
 \hline
N2 ABC (1) & $a_1=8.0232821323763118962115911006527082023129\times10^{-2}$ & 3.51\\ 
    & $a_2=6.6006740223496715389313719694385753827201\times10^{-2}$ & \\
N2 CAB (2) & $a_1=-6.9201301744275414774484171165883343266794\times10^{-2}$ & 2.91\\ 
     & $a_2=2.4031143347593460997280823526283232434071\times10^{-1}$ & \\
N2 ACB (1) & $a_1=2.6715152527177852877159258491775925759378\times10^{-1}$ & 1.18\\ 
     & $a_2=6.6006740223496715389313719694385753827201\times10^{-2}$ & \\
N2 BAC (2) & $a_1=4.5504624774591050429019276281136041301731\times10^{-2}$ & 1.06\\ 
     & $a_2=1.5208328361334726621353294430175796767150\times10^{-1}$ & \\
 \hline
N3 ABC (1) & $a_1=1.3174008291685690570191318391432506966565\times10^{-1}$ & 2.77\\ 
     & $b_1=2.5001213925191940517879462955880090794685\times10^{-1}$ & \\
N3 BAC (1) & $a_1=2.3903848575720093321093271828856414366272\times10^{-2}$ & 2.96\\ 
     & $b_1=4.2282680933338933433781797072334937140324\times10^{-1}$ & \\
 \hline
N4 BCA (1) & $a_1=2.2828507108154095724392719928157650301331\times10^{-1}$ & 3.22\\ 
     & $b_1=2.2825872461435056924300995377222198669807\times10^{-1}$ & \\
 \hline
N5 CAB (2) & $a_1=-6.2720924052603008551218820536171143473233\times10^{-2}$ & 1.59\\ 
     & $c_1=1.7666303579793115034995233223175505593095\times10^{-1}$ & \\
N5 ACB (1) & $a_1=2.2739584699362931383103719281935592654238\times10^{-1}$ & 2.84\\ 
     & $c_1=2.4520662064421018140607329552267381194383\times10^{-1}$ & \\
  N5 BAC (1) & $a_1=5.1047890551914167341876030732222399901020\times10^{-2}$ & 2.67\\ 
     & $c_1=2.2825872461435056924300995377222198669807\times10^{-1}$ & \\
 \hline
  N6 ABC (1) & $a_1=1.6014345007745294110506760786075550961823\times10^{-1}$ & 2.11\\ 
 & $b_1=3.3983727648480088011200434314807462997537\times10^{-1}$ & \\
    N6 ABC (2) & $a_1=3.4036466230135420614475440479339320092178\times10^{-1}$ & 2.19\\ 
& $b_1=1.6016272351519911988799565685192537002472\times10^{-1}$ & \\
     N6 BAC (1) & $a_1=6.6786520394832546068432401466879049634959\times10^{-2}$ & 3.27\\ 
   & $b_1=4.3305225085804317378510080225902343232508\times10^{-1}$ & \\
\hline
\end{tabular}
\caption{\label{table:sol_water}Coefficients of the best 4th-order integrators N for the water molecule.}
\end{table}

After the spherical top, we will test our integrators on an asymmetric body, the water molecule, which is a standard model on which the integrators of the free rigid body are tested notably in molecular dynamics \citep{dullweber1997,fasso2003,hairervilmart2006,omelyan2007}.
\cite{fasso2003} and \cite{hairervilmart2006} used the moments of inertia ($I_1=0.345$, $I_2=0.653$, $I_3=1$).
This does not correspond to a physical body for which the sum of two moments of inertia must be superior or equal to the third.
We prefer use the values ($I_{1}={10220}/{29376}\simeq0.348$, $I_{2}={19187}/{29376}\simeq0.653$, $I_{3}=1$) \citep{eisenbergkauzmann1969}.

The water molecule has different moments of inertia and it is then necessary to consider the six possible permutations.
For the water molecule, the integrators N have in total 90 real solutions for the coefficients $a_i$, $b_i$, $c_i$.
We can first discriminate the different solutions by comparing the analytical remainder estimated by the method of section \ref{sec:construction_analytical}.
The integrators, which present the smallest analytical remainders, are indicated in table \ref{table:sol_water}.
Their analytical remainders are very close and to determine precisely the best integrator, we need to compare them numerically. 

\begin{figure}
\centering
\input{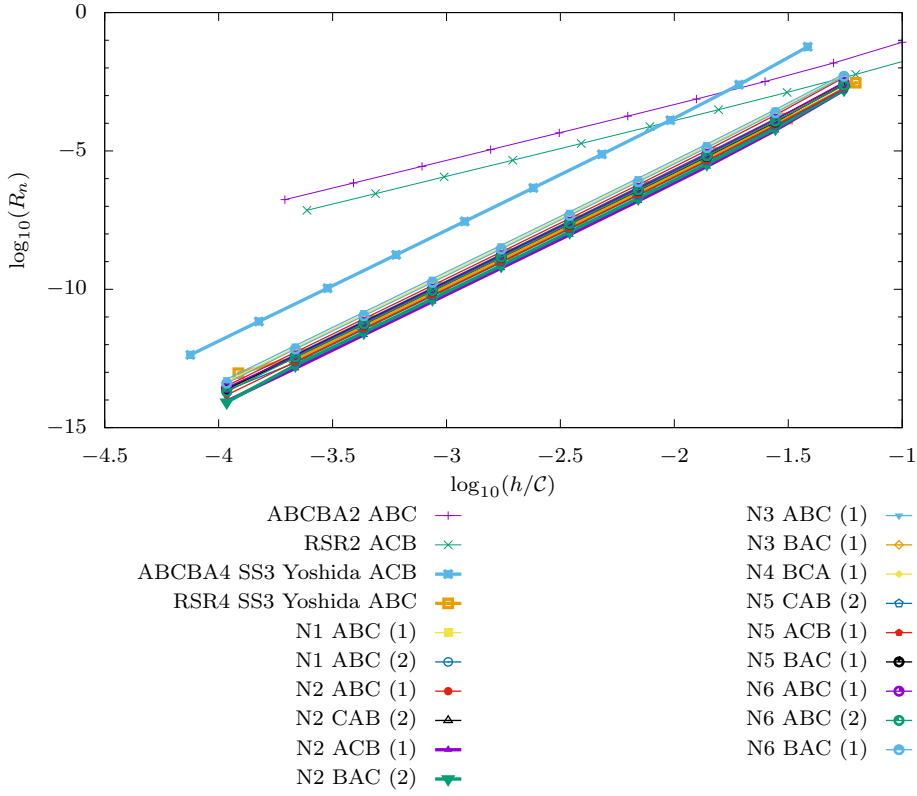} 
\caption{\label{fig:eau}Comparison of the numerical remainders $R_n$ for the water molecule between the integrators of the table \ref{table:sol_water} and the classical leapfrog scheme (ABCBA 2, RSR 2) and the Yoshida's scheme (ABCBA 4, RSR 4) obtained with the best permutation.}
\end{figure}
\begin{figure}
\centering
\input{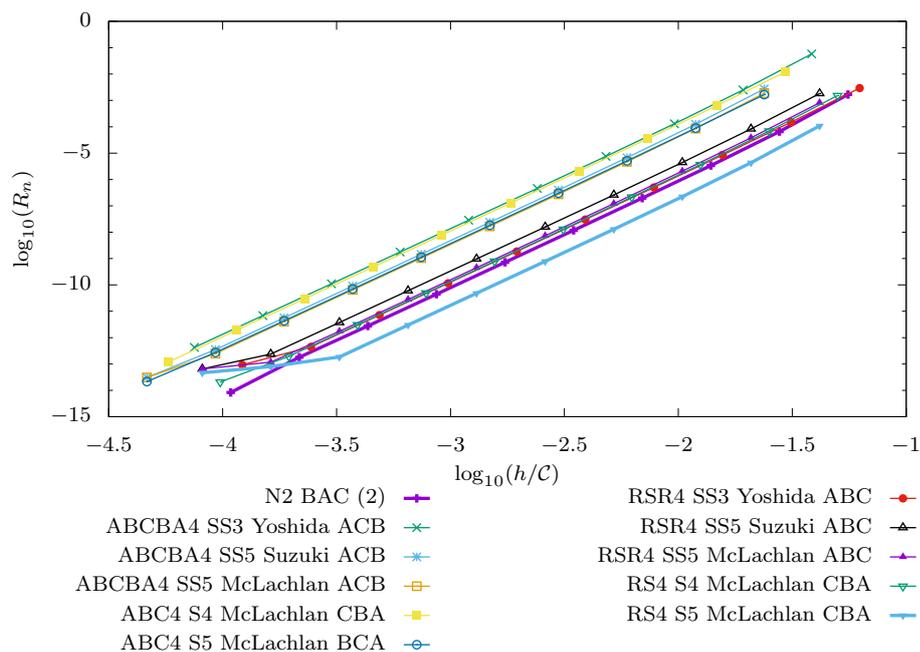} 
\caption{\label{fig:eau2}Comparison of the numerical remainders $R_n$ for the water molecule between the integrator N2 BAC (2) and the integrators obtained by \cite{yoshida1990}, \cite{suzuki1990} and \cite{mclachlan1995}.}
\end{figure}

We have represented on figure \ref{fig:eau} the obtained numerical remainders for the integrators of the table \ref{table:sol_water}, the classical leapfrog scheme and the Yoshida's integrator with the two splittings for their best permutations.

As predicted by the analytical remainder on table \ref{table:sol_water}, the numerical tests allow us to deduce that the best integrators for the water molecule are N2 BAC (2) and N2 ACB (1).
For the smallest step size, the best integrator is N2 BAC (2) and for the others N2 ACB (1).
We consider then that the best integrator is N2 BAC (2) for the water molecule.
It is respectively about 170 and 1.6 times more accurate than the Yoshida's integrators ABCBA 4 and RSR 4 obtained with the more accurate permutation of moments of inertia.
The best scheme for the water molecule $\mathcal{S}_{N2\,BAC\,(2)}\left(h\right)$ is given by
\begin{equation}
e^{a_{1}hL_{B}}e^{b_{1}hL_{A}}e^{a_{2}hL_{B}}e^{c_{1}hL_{C}}e^{a_{3}hL_{B}}e^{c_{1}hL_{C}}e^{a_{2}hL_{B}}e^{b_{1}hL_{A}}e^{a_1hL_{B}},\label{eq:scheme_2BAC(2)}
\end{equation}
with the coefficients
\begin{equation} 
\begin{array}{llll}  
 & a_{1} & = & 4.5504624774591050429019276281136041301731\times10^{-2} \\
 & b_{1} & = & \frac{1}{2} \\
 & a_{2} & = & 1.5208328361334726621353294430175796767150\times10^{-1} \\
 & c_{1} & = & \frac{1}{2} \\
 & a_{3} & = & 1-2(a_1+a_2).
\end{array}
\end{equation}
We notice that all the coefficients are positive.
This is not possible in a classical 4th-order symplectic integrator \citep{sheng1989,suzuki1991}.

It is possible to obtain 4th-order schemes more accurate than the Yoshida's one.
We then compare to the 4th-order schemes of \cite{suzuki1990} and \cite{mclachlan1995} and determine the best permutation for all these schemes. 
On figure \ref{fig:eau2}, we compare the numerical remainders of these integrators and deduce that the best is RS4 S5 McLachlan CBA.
On figure \ref{fig:eau2}, we also compare with the integrator N2 BAC (2) and conclude that RS4 S5 McLachlan CBA is about 4.7 times more accurate than N2 BAC (2).

\begin{figure}
\centering
\input{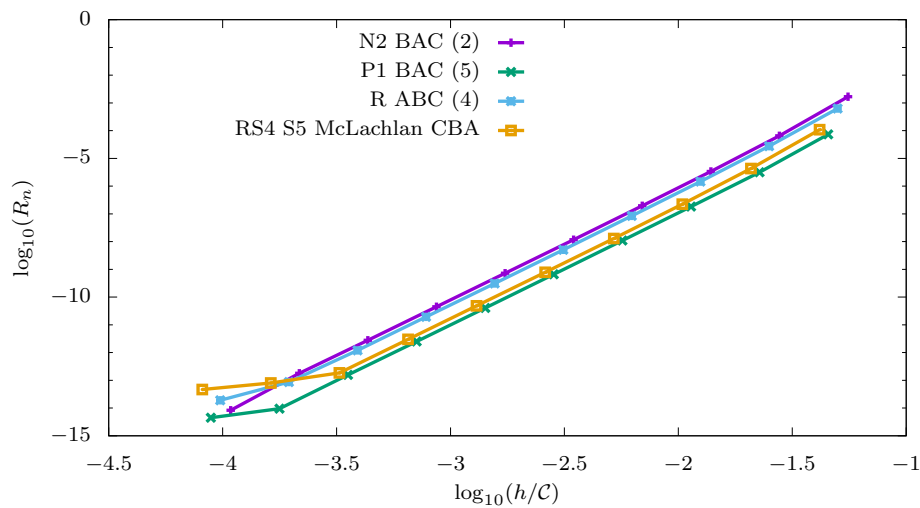} 
\caption{\label{fig:eau3} Comparison of the numerical remainders $R_n$ for the water molecule between the integrators N2 BAC (2), P1 BAC (5), R ABC (4) and RS4 S5 McLachlan CBA.}
\end{figure}

For the water molecule, we see that the integrators N are not more efficient than the existing integrators. Therefore we look if adding a parameter to minimize the 5th-order remainder (section \ref{sec:construction_additional}) allows us to obtain better integrators.
We start with the integrator R.
The six permutations allow us to obtain 46 solutions.
We determine numerically the best integrator, which is R ABC (4) and is 1.5 times more accurate than N2 BAC (2) (figure \ref{fig:eau3}).
We then consider the fifteen integrators P, which have in total 724 solutions for the water molecule.
We determine numerically the best integrator among the ones which present the smallest analytical remainders.
The best one is P1 BAC (5), which is about 8 times more accurate than N2 BAC (2) and about 1.7 times more precise than RS4 S5 McLachlan CBA (figure \ref{fig:eau3}). The scheme $\mathcal{S}_{P1\,BAC\,(5)}\left(h\right)$ is given by
\begin{equation}
e^{a_{1}hL_{B}}e^{b_{1}hL_{A}}e^{a_{2}hL_{B}}e^{b_{2}hL_{A}}e^{a_{3}hL_{B}}e^{c_{1}hL_{C}}e^{a_{3}hL_{B}}e^{b_{2}hL_{A}}e^{a_{2}hL_{B}}e^{b_{1}hL_{A}}e^{a_{1}hL_{B}},\label{eq:scheme_P1BAC(5)}
\end{equation}
with the coefficients
\begin{equation} 
\begin{array}{ccl}  
a_{1}&=&2.6576137190217391781483465189759344419197\times10^{-2}\\
b_{1}&=&2.8352180398306075206301328654179005782548\times10^{-1}\\
a_{2}&=&2.7103966011355754480520833151108230977015\times10^{-1}\\
b_{2}&=&\frac{1}{2}-b_1\\
a_{3}&=&\frac{1}{2}-(a_1+a_2)\\
c_{1}&=&1.
\end{array}
\end{equation}
It corresponds to a modest decreasing of the computation time of $12\%$.

Therefore, the dedicated integrators N, P and R for the water molecule do not allow us to decrease significantly the computation time with respect to the existing reference integrators.

\section{Comparison for the set of the rigid bodies\label{sec:comparison}}

In this part, we compare for each rigid body the obtained dedicated integrators to the usual ones.
However, we do not reproduce the numerical test which we have performed for the water molecule for each physical rigid body.
We have seen in section \ref{sec:numerical_sphere} that the evaluation of the analytical remainder allows us to estimate faster the best integrator for a rigid body.
We use then in this part this evaluation of the remainder to compare the dedicated integrators to the usual ones.

\subsection{Method}

\begin{figure}
\centering
\input{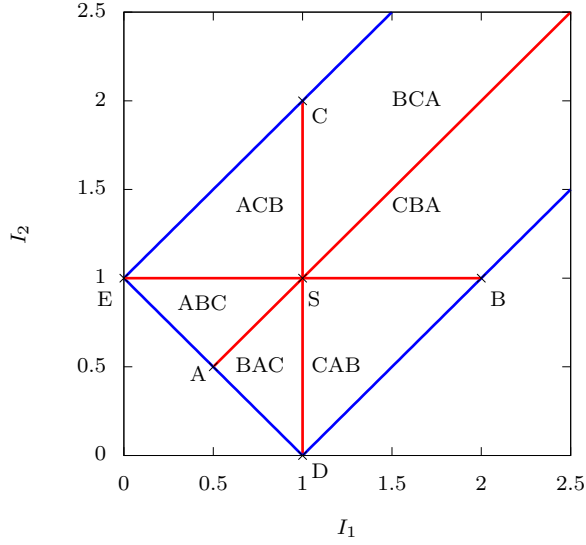} 
\caption{\label{fig:def_espace}
Set of the moments of inertia of the rigid bodies $\mathcal{E}$. The red lines correspond to the symmetric tops with two equal moments of inertia and delimit each permutation. The blue lines ($I_1+I_2=1$, $1+I_1=I_2$, $1+I_2=I_1$) correspond to the flat body and delimit the set $\mathcal{E}$. S corresponds to the spherical top and A, B, C, D, E to the flat symmetric tops.}
\end{figure}

We follow here the method of \cite{fasso2003}.
We consider a rigid body of moments of inertia ($I_1$, $I_2$, $I_3$), where the moments of inertia are normalized to obtain $I_3=1$. By definition \citep[e.g.][]{fasso2003}, the sum of any two moments of inertia must be superior or equal to the third one. Therefore $I_1$ and $I_2$ must verify 
\begin{equation}
I_1+I_2\geq 1, \qquad 1+I_1\geq I_2, \qquad 1+I_2\geq I_1.
\end{equation}
This allows one to define a set $\mathcal{E}$ represented on the figure \ref{fig:def_espace}. $\mathcal{E}$ is not bounded and $I_1$ and $I_2$ can go to infinity. However several points of $\mathcal{E}$ represent the same rigid body. For instance we consider the body ($I_1$, $I_2$, $1$) with $I_1<I_2<1$. This body can be represented in $\mathcal{E}$ by the six points ($I_1$, $I_2$), ($I_2/I_1$, $1/I_1$), ($1/I_2$, $I_1/I_2$), ($I_1/I_2$, $1/I_2$), ($1/I_1$, $I_2/I_1$),  ($I_2$, $I_1$) associated with the respective permutations of moments of inertia $ABC$, $BCA$, $CAB$, $ACB$, $CBA$, $BAC$.

An asymmetric body is then represented in the set $\mathcal{E}$ by six points corresponding to the six permutations.
As \cite{fasso2003}, we can restrict the study to the triangle ESA of the set $\mathcal{E}$ in figure \ref{fig:def_espace} provided we consider the six permutations for each point of this triangle.

\begin{figure}
\centering
\input{figures/tex_figure7a.tex} 
\input{figures/tex_figure7b.tex} 
\caption{\label{fig:XY}Best integrators N (a) and associated permutations (b) for a rigid body of moments of inertia ($I_1<I_2<1$), where each integrator and permutation are associated with a color.}
\end{figure}
\begin{figure}
\centering
\input{figures/tex_figure8.tex} 
\caption{\label{fig:XY_comp}Ratio of the remainder of the integrator RS4 S5 McLachlan obtained with the best permutation, $R_M$, on the one of the best integrator N, $R_N$, for a rigid body of moments of inertia ($I_1<I_2< 1$). The color scale indicates $\log_{10}(R_M/R_N)$.}
\end{figure}

We sample the triangle ESA with a grid of $N=40401$ points.
We do not consider the points of this grid which correspond to symmetric tops, which have a trivial solution.
For each point and for each permutation, we estimate the analytical remainder as explained in section \ref{sec:construction_analytical} for each dedicated integrator and for the integrator RS4 S5 McLachlan which is the best reference integrator for the water molecule.
We first determine the best dedicated integrator, which has the smaller analytical remainder, and its associated permutation.
Then we compare its analytical remainder to the one of the integrator RS4 S5 McLachlan for its best permutation.

\subsection{Integrator N}

We determine the best dedicated integrator N and the associated permutation in figure \ref{fig:XY} for the set of the rigid bodies.
If we know the moments of inertia of a body, we can then determine easily its best integrator N with the figure \ref{fig:XY}.

For each point of the triangular grid, we divide the analytical remainder of the integrator RS4 S5 McLachlan for the best permutation by the one of the best integrator N.
Along section \ref{sec:numerical_method}, RS4 S5 McLachlan has a cost of $12\mathcal{C}_0$ and an integrator N a cost of $9\mathcal{C}_0$.
We take into account the cost of these integrators of order 4 by multiplying the ratio by $(12/9)^4$.
The values of the ratio of these two remainders on the grid are in figure \ref{fig:XY_comp}.
If the ratio is larger than 1, the integrators N are better than RS4 S5 McLachlan.
We observe that the integrators N are better than the reference scheme RS4 S5 McLachlan only for some bodies of the triangular grid.
RS4 S5 McLachlan is especially more accurate for the ones close to the symmetric tops.  
Therefore the integrators N, which we have built, are in general less efficient than the existing integrators.

\begin{figure}
\centering
\input{figures/tex_figure9a.tex} 
\input{figures/tex_figure9b.tex} 
\caption{\label{fig:XYparam}Best integrators P (a) and associated permutations (b) for a rigid body of moments of inertia ($I_1<I_2<1$), where each integrator and permutation are associated with a color.}
\end{figure}
\begin{figure}
\centering
\input{figures/tex_figure10.tex} 
\caption{\label{fig:XYparam_comp}Ratio of the remainder of the integrator RS4 S5 McLachlan obtained with the best permutation, $R_M$, on the one of the best integrator P, $R_P$, for a rigid body of moments of inertia ($I_1<I_2< 1$). The color scale indicates $\log_{10}(R_M/R_P)$.}
\end{figure}

\subsection{Integrator P}

As made for the integrators N, we determine the best dedicated integrator P and the associated permutation in figure \ref{fig:XYparam} for the set of the rigid bodies. In figure \ref{fig:XYparam}, one can notice isolated points. They correspond to values for which a singularity occurs during the automatic resolution of the system of equations. We have not analyzed more precisely these solutions.

We represent in figure \ref{fig:XYparam_comp} the ratio of the analytical remainder of the integrator RS4 S5 McLachlan for the best permutation by the one of the best integrator P.
Along section \ref{sec:numerical_method}, an integrator P has a cost of $11\mathcal{C}_0$ and we have then multiplied the ratio by $(12/11)^4$.
The values of the ratio of these two remainders on the grid are in figure \ref{fig:XYparam_comp}.
We observe that the integrators P are better than the reference scheme RS4 S5 McLachlan only for very asymmetric bodies.

\section{Conclusion}

We used the properties of the Lie algebra of the angular momentum to build symplectic integrators dedicated to the Hamiltonian of the free rigid body.
The relation $\lbrace G_{i},G_{j}\rbrace = \epsilon_{ijk} G_{k}$ between the components of the angular momentum simplifies the expression of the remainders of a symplectic integrator of the rigid body.
These remainders depend on the moments of inertia of the integrated body.
By introducing a dependence of the coefficients of the integrators on the moments of inertia, we can cancel the 3rd-order remainder to construct symmetric 4th-order integrators dedicated to the rigid bodies.
For the splitting in three parts (splitting ABC), it allows us to obtain symplectic 4th-order integrators for the free rigid body with fewer stages than for the general case.
On the opposite, this reduction does not occur for the integrators obtained with the splitting in two parts (splitting RS).
During our analysis of the splitting RS, we have noted a commutation that allows us to decrease the computation time of the RS integrators, which has not been previously noticed as far as we know.

We performed extensive numerical tests on the water molecule which is a classical body to test integrators of rigid bodies.
We first test the obtained dedicated 4th-order integrators with the minimal number of stages (integrators N).
These integrators for the water molecule are not more efficient than the existing ones.
We then consider the integrators which we have obtained by adding a free parameter to minimize the 5th-order remainders (integrators R and P).
Deceptively, these integrators for the water molecule are only slightly more accurate than the existing reference integrators.

By sampling the set of the moments of inertia of the rigid bodies, we determine for each existing body the best 4th-order integrators N and P and the associated permutation by estimating the analytical 5th-order remainder.
We then compare the best new integrator for each body to the best existing integrator for the water molecule.
The integrators N have not better performances than the existing ones while the integrators P can be better for very asymmetric bodies.

Here we restricted ourselves to the simpler 4th-order integrators to obtain simpler schemes.
It should be still possible to obtain better schemes by considering the addition of two free parameters.
However, the coefficients are then more difficult to obtain and the schemes more complicate than the ones, which have been obtained here.
It is also possible to construct 6th-order integrators but the system of equations to solve to obtain the coefficients becomes more difficult to solve.

\appendix
\section{\label{AppendixA}Solutions for the 4th-order integrators N}

For each 4th-order integrator N of section \ref{sec:construction_N}, the values of the coefficients $a_i$, $b_i$, $c_i$ are given by the following equations for moments of inertia determined by the values of $1+x=I_1/I_2$ and $1+y=I_1/I_3$. 

\subsection*{N1: ABABCBABA}

\begin{eqnarray}
f_0+a_1f_1+a_1^{2}f_2+a_1^{3}f_3 & =& 0 \\
g_0+b_1g_1+a_1g_2 & = & 0 \nonumber
\end{eqnarray}
\begin{eqnarray}
f_0 &=& -1-3y-3y^{2}-2xy^{2}-3x^{2}y-12x^{2}y^{2}-12x^{2}y^{3}-4x^{2}y^{4}  \nonumber \\
f_1 &=& 6+18y+24y^{2}-12xy+12xy^{2}-18x^{2}-78x^{2}y-72x^{2}y^{2}-24x^{2}y^{3}  \nonumber \\ 
f_2 &=& -12-36y-72y^{2}-48x-72xy-168xy^{2}-48xy^{3}-36x^{2}+12x^{2}y  \nonumber \\
f_3 &=& -24-72y-144xy-48xy^{2}+72x^{2}+24x^{2}y  \nonumber \\ 
g_0 &=& -1-y-4x-6xy-2xy^{2}-3x^{2}-5x^{2}y-2x^{2}y^{2}  \nonumber \\ 
g_1 &=& 4+6y+10x+18xy+4xy^{2}+6x^{2}+12x^{2}y+4x^{2}y^{2}  \nonumber \\ 
g_2 &=& -2-6y-12xy-4xy^{2}+6x^{2}+2x^{2}y \nonumber
\end{eqnarray}

\subsection*{N2: ABACACABA}

\begin{eqnarray}
f_0+a_1f_1+a_1^{2}f_2 &=&0 \\ 
g_0+a_2g_1+a_1g_2 &=&0 \nonumber
\end{eqnarray}
\begin{eqnarray}
f_0 &=& 1+3y-3y^{3}-8xy^{2}-12xy^{3}+x^{2}y-3x^{2}y^{2}-9x^{2}y^{3}+x^{2}y^{4}-4x^{3}y^{3}+x^{4}y^{2}  \nonumber \\ 
f_1 &=& -6-30y-18y^{2}+18y^{3}-48xy-48xy^{2}+48xy^{3}+6x^{2}-30x^{2}y-78x^{2}y^{2}  \nonumber \\ 
&& +30x^{2}y^{3}-48x^{3}y^{2}+12x^{4}y  \nonumber \\ 
f_2 &=& 36y+36y^{2}-36y^{3}-36x+36xy+180xy^{2}-36xy^{3}-72x^{2}-144x^{2}y  \nonumber \\ 
&& +144x^{2}y^{2}-144x^{3}y+36x^{4}  \nonumber \\ 
g_0 &=& -1-3y-3y^{2}-2xy^{2}+x^{2}y \nonumber \\ 
g_1 &=& 6+12y+6y^{2}  \nonumber \\ 
g_2 &=& 6y^{2}-12xy+6x^{2} \nonumber  
\end{eqnarray}

\subsection*{N3: ABACBCABA}

\begin{eqnarray}
f_0+a_1f_1+a_1^{2}f_2+a_1^{3}f_3 &=&0 \\ 
g_0+b_1g_1+a_1g_2+a_1^{2}g_3 &=&0 \nonumber
\end{eqnarray}
\begin{eqnarray}
f_0 &=& 1+3y-3y^{3}-4xy^{2}-6xy^{3}+x^{2}y^{4}  \nonumber \\ 
f_1 &=& -6-30y-18y^{2}+18y^{3}-24xy-24xy^{2}+24xy^{3}+12x^{2}y^{3}  \nonumber \\ 
f_2 &=& 60y+72y^{2}-36y^{3}-24x+36xy+144xy^{2}-12xy^{3}+48x^{2}y^{2}  \nonumber \\ 
f_3 &=& 24-72y^{2}+72x+144xy-24xy^{2}+48x^{2}y  \nonumber \\ 
g_0 &=& -3-18y-39y^{2}-36y^{3}-12y^{4}-2x-15xy-39xy^{2}-41xy^{3}-15xy^{4}  \nonumber \\ 
&& -2x^{2}y^{2}-3x^{2}y^{3}+x^{2}y^{5}  \nonumber \\ 
g_1 &=& 2+12y+30y^{2}+36y^{3}+18y^{4}+2x+12xy+34xy^{2}+48xy^{3}+30xy^{4}  \nonumber \\ 
&& +4x^{2}y^{2}+12x^{2}y^{3}+14x^{2}y^{4}+2x^{3}y^{4}  \nonumber \\ 
g_2 &=& 2+30y+84y^{2}+78y^{3}+18y^{4}-6x+6xy+78xy^{2}+102xy^{3}+24xy^{4}  \nonumber \\ 
&& +4x^{2}y+30x^{2}y^{2}+48x^{2}y^{3}+10x^{2}y^{4}+4x^{3}y^{3}  \nonumber \\ 
g_3 &=& 12+24y-24y^{2}-72y^{3}-36y^{4}+36x+144xy+168xy^{2}+48xy^{3}-12xy^{4}  \nonumber \\ 
&& +24x^{2}y+48x^{2}y^{2}+24x^{2}y^{3} \nonumber 
\end{eqnarray}

\subsection*{N4: ABCABACBA}

\begin{eqnarray}
f_0+a_1f_1+a_1^{2}f_2+a_1^{3}f_3 &=&0 \\ 
g_0+b_1g_1+a_1g_2 &=&0 \nonumber
\end{eqnarray}
\begin{eqnarray}
f_0 &=& 1+3y-3y^{3}-4xy^{2}-6xy^{3}+x^{2}y^{4}  \nonumber \\ 
f_1 &=& -6-30y-18y^{2}+18y^{3}-24xy-24xy^{2}+24xy^{3}+12x^{2}y^{3}  \nonumber \\ 
f_2 &=& 60y+72y^{2}-36y^{3}-24x+36xy+144xy^{2}-12xy^{3}+48x^{2}y^{2}  \nonumber \\ 
f_3 &=& 24-72y^{2}+72x+144xy-24xy^{2}+48x^{2}y  \nonumber \\ 
g_0 &=& -1-y-2x-3xy-xy^{2}  \nonumber \\ 
g_1 &=& 2+6y+6y^{2}+2x+6xy+8xy^{2}+2x^{2}y^{2}  \nonumber \\ 
g_2 &=& 2-6y^{2}+6x+12xy-2xy^{2}+4x^{2}y \nonumber 
\end{eqnarray}

\subsection*{N5: ABCACACBA}

\begin{eqnarray}
f_0+a_1f_1+a_1^{2}f_2+a_1^{3}f_3 &=&0  \\ 
g_0+c_1g_1+a_1g_2+a_1^{2}g_3 &=&0 \nonumber
\end{eqnarray}
\begin{eqnarray}
f_0 &=& 1+3y-6xy^{2}-x^{2}y-6x^{2}y^{2}+x^{4}y^{2}  \nonumber \\ 
f_1 &=& -6-30y-36xy+36xy^{2}-6x^{2}-42x^{2}y+24x^{2}y^{2}+12x^{4}y  \nonumber \\ 
f_2 &=& 84y-48x+144xy-72xy^{2}-36x^{2}+180x^{2}y-24x^{2}y^{2}+24x^{3}y+36x^{4}  \nonumber \\ 
f_3 &=& 24-72y+144x-144xy+48xy^{2}+216x^{2}-168x^{2}y+144x^{3}  \nonumber \\ 
g_0 &=& 1+4y+x+9xy-8xy^{2}-3x^{2}+4x^{2}y-25x^{2}y^{2}-6x^{3}-7x^{3}y-30x^{3}y^{2}-2x^{3}y^{3}\nonumber \\ 
&& -3x^{4}-9x^{4}y-12x^{4}y^{2}-2x^{4}y^{3}-6x^{5}y+x^{5}y^{2}-3x^{6}y  \nonumber \\ 
g_1 &=& 2+2y+12x+12xy+30x^{2}+34x^{2}y+4x^{2}y^{2}+36x^{3}+48x^{3}y+12x^{3}y^{2}+18x^{4}\nonumber \\ 
&& +30x^{4}y+14x^{4}y^{2}+2x^{4}y^{3}  \nonumber \\ 
g_2 &=& -4-30y-6x-114xy+28xy^{2}-6x^{2}-210x^{2}y+66x^{2}y^{2}-18x^{3}-210x^{3}y+48x^{3}y^{2}\nonumber \\ 
&& +4x^{3}y^{3}-36x^{4}-96x^{4}y-2x^{4}y^{2}-36x^{5}-18x^{6}  \nonumber \\ 
g_3 &=& -12+36y-96x+144xy-24xy^{2}-264x^{2}+264x^{2}y-48x^{2}y^{2}-360x^{3}+240x^{3}y\nonumber \\ 
&& -24x^{3}y^{2}-252x^{4}+84x^{4}y-72x^{5} \nonumber 
\end{eqnarray}

\subsection*{N6: ABCBABCBA}

\begin{eqnarray}
f_0+a_1f_1+a_1^{2}f_2+a_1^{3}f_3+a_1^{4}f_4 &=&0 \\ 
g_0+b_1g_1+a_1g_2+a_1^{2}g_3+a_1^{3}g_4 &=&0 \nonumber
\end{eqnarray}
\begin{eqnarray}
f_0 &=& 1+3y+3y^{2}-3y^{3}+8xy^{2}-3x^{2}y+3x^{2}y^{2}+3x^{2}y^{3}+x^{2}y^{4}  \nonumber \\ 
f_1 &=& -6-18y-42y^{2}+18y^{3}+48xy-48xy^{2}-18x^{2}+42x^{2}y+18x^{2}y^{2}+6x^{2}y^{3}  \nonumber \\ 
f_2 &=& 12+36y+180y^{2}-36y^{3}+12x-252xy+132xy^{2}+12xy^{3}+144x^{2}-48x^{2}y  \nonumber \\ 
f_3 &=& -288y^{2}+576xy-288x^{2}  \nonumber \\ 
f_4 &=& 144y^{2}-288xy+144x^{2}  \nonumber \\ 
g_0 &=& -2y-6y^{2}+4x+11xy-xy^{2}+xy^{3}+xy^{4}+3x^{2}+13x^{2}y+7x^{2}y^{2}+x^{2}y^{3}  \nonumber \\ 
g_1 &=& -4-10y-6y^{2}-6y^{3}-6y^{4}-10x-28xy-16xy^{2}-4xy^{3}-6xy^{4}-6x^{2}\nonumber \\ 
& & -18x^{2}y-10x^{2}y^{2}+2x^{2}y^{3}  \nonumber \\ 
g_2 &=& 4+14y+42y^{2}+6y^{3}+6y^{4}+2x-42xy+14xy^{2}-6xy^{3}+24x^{2}-4x^{2}y+4x^{2}y^{2}  \nonumber \\ 
g_3 &=& -84y^{2}-12y^{3}+168xy+24xy^{2}-84x^{2}-12x^{2}y  \nonumber \\ 
g_4 &=& 48y^{2}-96xy+48x^{2} \nonumber 
\end{eqnarray}

\subsection*{N7: ABCBCBCBA}

\begin{eqnarray}
f_0+b_1f_1+b_1^{2}f_2+b_1^{3}f_3 &=&0  \\ 
g_0+c_1g_1+b_1g_2 &=&0 \nonumber
\end{eqnarray}
\begin{eqnarray}
f_0 &=& -1-3y-3x-15xy-6xy^{2}-17x^{2}y-12x^{2}y^{2}+3x^{3}-3x^{3}y-6x^{3}y^{2}-x^{4}y^{2}  \nonumber \\ 
f_1 &=& 12+30y+42x+150xy+36xy^{2}+30x^{2}+222x^{2}y+84x^{2}y^{2}-18x^{3}+114x^{3}y \nonumber \\ 
&& +60x^{3}y^{2}-18x^{4}+12x^{4}y+12x^{4}y^{2}  \nonumber \\ 
f_2 &=& -48-84y-204x-420xy-72xy^{2}-300x^{2}-708x^{2}y-192x^{2}y^{2}-180x^{3}-492x^{3}y \nonumber \\
&& -168x^{3}y^{2}-36x^{4}-120x^{4}y-48x^{4}y^{2}  \nonumber \\ 
f_3 &=& 48+72y+216x+360xy+48xy^{2}+360x^{2}+648x^{2}y+144x^{2}y^{2}+264x^{3}+504x^{3}y\nonumber \\
&& +144x^{3}y^{2}+72x^{4}+144x^{4}y+48x^{4}y^{2}  \nonumber \\ 
g_0 &=& y-x+3xy+2xy^{2}-3x^{2}-2x^{2}y  \nonumber \\ 
g_1 &=& 2+2y+6x+6xy+6x^{2}+8x^{2}y+2x^{2}y^{2}  \nonumber \\ 
g_2 &=& -4-6y-10x-18xy-4xy^{2}-6x^{2}-12x^{2}y-4x^{2}y^{2} \nonumber
\end{eqnarray}


\section*{Conflict of interest}

The authors declare that they have no conflict of interest.

\bibliographystyle{spbasic}      
\bibliography{symrot}   

\end{document}